\documentclass[a4paper,11pt]{article}
\usepackage{jheppub} 
\usepackage{lineno}
\usepackage[normalem]{ulem}
\usepackage{cancel}
\usepackage{xcolor}
\usepackage{float}

\newcommand{\jpb}[1]{{\color{purple}\bf [JPB: #1]}}

\newcommand{\A}[1]{{\color{teal}\bf [Aoumeur: #1]}}

\newcommand{\pbg}[1]{{\color{orange}\bf [PBG: #1]}}

\newcommand{\tg}[1]{{\color{brown}\bf [TG~: #1]}} 

\newcommand{\rmd}{{\mathrm d}}
\newcommand{\rme}{{\mathrm e}}
\newcommand \beq{\begin{eqnarray}}
	\newcommand \eeq{\end{eqnarray}}
\def\x{{\boldsymbol x}}
\def\r{{\boldsymbol r}}
\def\s{{\boldsymbol s}}
\def\p{{\boldsymbol p}}
\def\q{{\boldsymbol q}}
\def\k{{\boldsymbol k}}
\def\z{{\boldsymbol z}}
\def\y{{\boldsymbol y}}
\def\J{{\boldsymbol J}}
\def\R{{\boldsymbol R}}
\def\P{{\boldsymbol P}}
\def\X{{\boldsymbol X}}
\def\bfq{{\boldsymbol q}}

\def\bra#1{\langle#1\vert}
\def\ket#1{\vert#1\rangle}
\newcommand{\nn}{\nonumber\\ }

\title{General Lindblad equation for quarkonium evolution in a quark-gluon plasma}

\author[a]{Aoumeur Daddi Hammou}
\author[b]{Jean-Paul Blaizot}
\author[a]{Pol Bernard Gossiaux}
\author[a]{Thierry Gousset}

\affiliation[a]{SUBATECH, IMT Atlantique, Nantes Université, CNRS/IN2P3,  Nantes, 44307, France}
\affiliation[b]{Institut de physique théorique, Université Paris-Saclay, CNRS, CEA, Gif-sur-Yvette 91191, France}

\emailAdd{gossiaux@subatech.in2p3.fr}

\abstract{Accurate modelling and understanding of quarkonium production in ultrarelativistic heavy ion collisions requires a formalism that preserves the quantum properties of microscopic $\bar{Q}Q$
	systems while treating the interaction of such pairs with the quark-gluon plasma (QGP). The open quantum system approach has recently emerged as one of the most fruitful schemes to meet such requirements. However, the quantum master equations obtained so far in this context are derived assuming a strict ordering between the QGP temperature $T$ and the $\bar{Q}Q$ energy gaps ($\Delta E$) of the  quarkonia bound states. This limits their predictive power since, as the QGP expands and cools down, the system traverse all regimes between the quantum Brownian motion (QBM) regime for $T\gtrsim \Delta E$ to the quantum optical (QO) regime  for $\Delta E\gtrsim T$. In this paper, we derive and present a more general non-abelian quantum master equation of the Lindblad type, which does not suffer from these limitations and thus allows to faithfully describe the quantum evolution of the $Q\bar{Q}$ pairs during the whole QGP-evolution. We also provide some illustration of the key quantities governing this equation.}

\begin{document}
	\maketitle
	\flushbottom

	\section{Introduction}
	\label{sec:intro}
	
	It is nowadays established that ultrarelativistic heavy ion  collisions (URHIC) lead to the creation of a new state of strongly interacting matter, known as the quark-gluon plasma (QGP), where the elementary constituents of usual hadrons, quarks and gluons, are deconfined. Understanding QGP properties is at the core of extensive dedicated theoretical and experimental studies~\cite{shuryak1978quark,Gyulassy:2004zy,Shuryak:2004cy,doi:10.1142/9533,PHENIX:2004vcz,STAR:2005gfr,ALICE:2022wpn,Collaboration2024Overview,Spousta2021Highlights}. 
	
	As the QGP droplets formed in URHIC are quite elusive, living for $\approx 10^{-22}\,{\rm s}$, ``measuring" QGP properties can only be achieved in an indirect way, resorting e.g. to so-called ``hard probes" created in the early stage of URHIC and interacting with the QGP during its evolution. Among the various hard probes of the QGP, the suppression of (the production of) heavy quarkonia constitutes one of the most important ones owing to its sensitivity to various mechanisms such as the modification of the strong force at short distance (colour screening) or the dissociation of bound states by inelastic process with QGP constituents \cite{Bhanot:1979vb}. 
	
	Since its suggestion as a QGP probe in 1986~\cite{matsui1986j}, quarkonia production in URHIC has been the subject of intensive theoretical and experimental research, resulting in fruitful cross-fertilisation between both perspectives, see e.g. \cite{andronic:2024oxz,Andronic:2025jbp} for recent reviews. Along the 40 last years, the theoretical modelling of quarkonia production in URHIC has evolved from considering either the effect of the Debye screening of the vacuum potential through simple criteria, e.g. in  \cite{Chu:1987sy} (where the local QGP temperature is compared to the so-called dissociation temperature of each bound state after its formation time) or the role of inelastic collisions with the QGP constituents~\cite{PhysRevD.39.232}, to semi-classical transport approaches \cite{Grandchamp:2003uw,Gossiaux:2004qw,Young:2008he,Zhao:2010nk,Liu:2009nb,Liu:2010ej,Nendzig:2012cu,Zhao:2017yan,Villar:2022sbv,Song:2023zma,Wu:2024gil} 
	of increasing sophistication including both effects as well as new ones such as $Q\bar{Q}$ recombination~\cite{Braun-Munzinger:2000csl,Thews:2000rj}. 
	
	An important factor in improving these real-time transport approaches is the steady progress achieved in lattice-QCD (lQCD) computations (see e.g. \cite{Bazavov:2023dci,Ali:2025iux}). Despite the fact that real-time simulations are still out of reach, the use of sophisticated inversion methods allows to extract valuable information pertaining to the in-QGP quarkonium dynamics, such as their spectral functions as well as the complex potential for a static $Q\bar{Q}$ pair, whose imaginary part represents~\cite{laine2007real,beraudo2008real} the width of the quarkonium peaks in the spectral function. Combined with effective theories such as potential non-relativistic QCD (pNRQCD)~\cite{brambilla2011thermal}, these lQCD inputs are highly effective in constraining the various models~\cite{Tang:2023tkm}.  
	
	Another mandatory factor to improve the faithfulness of these models is to be able to cope with the quantum nature of quarkonium states. Early attempts in this direction~\cite{Cleymans:1989is,Cugnon:1993ye}  were based on numerical solutions of time-dependent Schrödinger equations and could then only cope with the screening of the real-potential. However, as quarkonia propagate through the hot and dense QGP medium, they continuously interact with the surrounding plasma, leading to dissipation and quantum decoherence. In such situations, the subsystem of interest cannot, in general, be regarded as isolated, making the open quantum systems (OQS) formalism~\cite{breuer2002theory,rivas2012open,joos2013decoherence} a natural framework for describing its real-time evolution. Over the past decade, the open quantum systems formalism has indeed emerged as a powerful first-principles framework for describing their in-medium evolution (see~\cite{akamatsu2022quarkonium,Yao:2021lus} for recent reviews) as well as for establishing a direct bridge between the underlying quantum dynamics and the semiclassical transport equations~
	\cite{blaizot2018quantum,Daddi-Hammou:2025hdz,yao2021semiclassical-transport} which are often used as a proxy to obtain numerical solutions.
	
	Within the OQS framework, the real-time evolution of quarkonia can be described using several complementary approaches, including path integral and influence functional methods (see e.g. \cite{beraudo2010path,Blaizot:2015hya,DeBoni:2017ocl}) and quantum master equations (QME) governing the dynamics of the (reduced) density matrix (see e.g.~\cite{brambilla2017quarkonium,brambilla2018heavy,blaizot2018quantum,blaizot2018approach,Yao:2021lus,Miura:2022arv,Brambilla:2022ynh,Delorme:2024rdo}), while Schrödinger-Langevin equations~\cite{katz2016schrodinger}, although more phenomenological, may offer an interesting alternative as their numerical computation is less demanding.   
	
	Most of those QME are established in the Markovian limit, assuming that the quarkonium relaxation time $\tau_R$ is large compared to the QGP autocorrelation time $\tau_E$. In this Markovian limit, complete positivity requires the resulting evolution equation to take the Lindblad form~\cite{lindblad1976generators,Gorini:1975nb}. Moreover, Lindblad QME can be solved resorting to by-now standard quantum diffusion methods established in the early 1990s~\cite{gisin1992quantum,Molmer:93}, where the deterministic dynamics of the density matrix is shown to be equivalent to an ensemble average of quantum states evolving according to stochastic equations. This explains why most of the approaches aiming for an exact numerical solution seek to formulate their QME in the Lindbladian form.\footnote{While direct solution of the original QME is also feasible, at the price of considering 1D systems~\cite{Delorme:2024rdo} or truncating the number of spherical harmonics in 3D space~\cite{brambilla2018heavy}.}  
	To achieve this goal, further assumptions regarding the relative magnitude of the plasma temperature $T$ and the characteristic quarkonium energy gaps, $\Delta E$ are usually performed~\cite{akamatsu2022quarkonium,Yao:2021lus}. Consequently, distinct Lindblad equations describe the quantum Brownian motion (QBM) regime, $T\gtrsim\Delta E$, and the quantum optical (QO) regime, $\Delta E\gtrsim T$, with each equation being restricted to its corresponding domain of validity.
	As the QGP expands and cools down, however, the relative magnitude of $T$ and $\Delta E$ changes continuously, and both of these two regimes apply during the full QGP-history.\footnote{This has recently led some authors to consider $\Delta E/T$ corrections in the QBM regime~\cite{Brambilla:2022ynh,Delorme:2024rdo} to extend its applicability to lower values of $T$.} This caveat motivates the development of a unified description that remains valid throughout the entire evolution of the QGP. 
	
	In this work, we address this issue and formulate the dynamics of the quarkonium-QGP system resorting to the universal Lindblad equation (ULE) introduced recently by Nathan and Rudner~\cite{nathan2020universal}. Similarly to the Redfield equation from which it is derived, the ULE relies solely on the Born-Markov approximation and does not further assume any hierarchy between the QGP temperature $T$ and the characteristic quarkonium energy gaps $\Delta E$, thereby overcoming the limitations of previous Lindblad formulations. The resulting coupled singlet-octet dynamics stemming from these equations provide a unified, completely positive description of in-medium quarkonium dynamics that encompasses both the quantum Brownian and the quantum optical regimes within a single theoretical framework. 
	
	To illustrate the physical content of the resulting QCD-universal Lindblad equations, we present several illustrative results. In particular, we investigate the temperature dependence of the singlet decay rate induced by singlet-octet transitions and analyse the spectral density in the singlet channel. These calculations provide a first illustration of the physical content of the universal Lindblad framework and show that it represents a unified and physically consistent description of representative in-medium quarkonium observables on a temperature range covering both the QO and the QBM regimes. A complete numerical solution of the coupled singlet-octet universal Lindblad equations is left for future work. 
	
	This paper is organised as follows. In Sec.~\ref{NRQCD-ULEb}, we derive the coupled singlet-octet universal Lindblad equations for a $Q\bar{Q}$ pair. In Sec.~\ref{sub_sec_illustrations}, we present illustrative results projecting the ULE on the eigenstates of the $Q\bar{Q}$ hamiltonian and use the evolution equations obtained in such a way to make contact with the Redfield QME, in particular in the QO regime, while we  discuss the connection with equations commonly used in the QBM regime in Sec.~\ref{sec_contact}. Finally, in Sec.~\ref{section:Conclusions}, we summarise our main results and outline directions for future work.
	
	\section{A QCD universal Lindblad equation}
	\label{NRQCD-ULEb}
	
	We consider a single heavy quark-antiquark pair immersed in a plasma of light quarks and gluons, in thermal equilibrium at a temperature $T$ much smaller than the mass $M$ of the heavy quark. Relying on the condition $M\gg T$ we treat the heavy quark and antiquark as  non-relativistic particles, and we neglect their magnetic interactions (among themselves, and with the plasma constituents). We assume then that the whole system can be described by the following Hamiltonian
	\begin{equation}\label{hamiltonian}
		H=H_{\rm pl}+H_{\scriptscriptstyle Q}+H_1\,,
	\end{equation}
	where $H_{\rm pl}$ is the QCD Hamiltonian governing the dynamics of the plasma in the absence of the heavy quarks,  while $H_{\scriptscriptstyle Q}$ is the hamiltonian of the heavy quark-antiquark pair in the absence of the plasma. Note that $H_{\scriptscriptstyle Q}$ includes the direct colour static Coulomb-like interaction between the quark and the antiquark. Its explicit form will be given later when needed. The last term in Eq.~(\ref{hamiltonian}) is the interaction between the plasma and the heavy quarks. In the Coulomb gauge, it is of the form\footnote{Throughout this paper, we use the shorthand notation  $\int_\x\equiv \int \rmd^3 \x$ for the spatial integrals, and $\int_\p\equiv \int \frac{\rmd^3\p}{(2\pi)^3}$ for momentum integrals.}
	\beq\label{H1}
	H_1=-g\int_\x A_0^{\scriptscriptstyle A}(\x) n^{\scriptscriptstyle A}(\x),
	\eeq
	where $A_0^{\scriptscriptstyle A}$ denotes the (colour) Coulomb field created by the plasma particles, while $n^{\scriptscriptstyle A}$ denotes the colour charge density of the heavy particles, in first quantisation \cite{blaizot2018quantum}:
	\beq\label{colourdensity}
	n^{\scriptscriptstyle A}(\x)=\delta(\x-\hat \r)\, t^{\scriptscriptstyle A}\otimes \mathbb{I} -\mathbb{I}\otimes\delta(\x-\hat\r) \, \tilde t^{\scriptscriptstyle A}.
	\eeq
	Here $\hat \r$ denotes the position operator\footnote{ We occasionally put a hat on operators whenever confusion may arise from not doing so.}, and  the two components of the tensor product refer respectively to the heavy quark (for the first component) and to the heavy antiquark (for the second component).
	In Eq.~(\ref{colourdensity}), $t^{\scriptscriptstyle A}$ is a colour matrix in the fundamental representation of SU(3),  coupling the heavy quark to  the gluon field. The coupling of the heavy antiquark to the gluon is described by $-\tilde t^{\scriptscriptstyle A}$, with $\tilde t^{\scriptscriptstyle A}$ the transpose of $t^{\scriptscriptstyle A}$. We use capital letter upperscripts, as in  $t^{\scriptscriptstyle A}$, to denote the $(N_c^2-1)$ indices of the adjoint representation. 
	
	\subsection{The Nathan-Rudner  equation}
	\label{sec:NathanRudner}
	
	The goal of this section is to review the derivation of the so-called Universal Lindblad Equation (ULE) proposed by Nathan and Rudner~\cite{nathan2020universal}, with in mind its application to  the evolution of a heavy quark-antiquark pair in a quark-gluon plasma. Our starting point is the equation for the reduced density matrix derived in~\cite{blaizot2018approach}, and which reads
	\begin{equation}
		\begin{split}
			\frac{\rmd \bar\rho\left(t\right)}{\rmd t} =- & \left(\int_{t_{0}}^{t}dt^{\prime}\int_{\boldsymbol{x}\boldsymbol{x}^{\prime}}\left[\bar n^{\scriptscriptstyle A}\left(t,\boldsymbol{x}\right),\bar n^{\scriptscriptstyle A}\left(t^{\prime},\boldsymbol{x}^{\prime}\right)\bar \rho \left(t^{\prime}\right)\right]\Delta^{>}\left(t-t^{\prime},\boldsymbol{x}-\boldsymbol{x}^{\prime}\right)\right.\\
			&\left.+\int_{t_{0}}^{t}dt^{\prime}\int_{\boldsymbol{x}\boldsymbol{x}^{\prime}}\left[\bar \rho\left(t^{\prime}\right)\bar n^{\scriptscriptstyle A}\left(t^{\prime},\boldsymbol{x}^{\prime}\right),\bar n^{\scriptscriptstyle A}\left(t,\boldsymbol{x}\right)\right]\Delta^{<}\left(t-t^{\prime},\boldsymbol{x}-\boldsymbol{x}^{\prime}\right)\right).
		\end{split}
		\label{eq:be}
	\end{equation}
	In this equation, $\bar \rho$  is the reduced density matrix of the heavy quark-antiquark pair,  and $\bar n_{\boldsymbol{x}}^{\scriptscriptstyle A}$ is the colour charge density (\ref{colourdensity}). A sum over   repeated colour indices is implied throughout. Both $\bar \rho$ and $\bar n^{\scriptscriptstyle A}$ are written in the interaction representation (indicated by a bar), the  unperturbed hamiltonian  being $H_{\scriptscriptstyle Q}+H_{\rm pl}$.  Thus, for instance,  $\bar n^{\scriptscriptstyle A}(t,\boldsymbol{x}) =\rme^{iH_{\scriptscriptstyle Q} (t-t_0)} n^{\scriptscriptstyle A}(\boldsymbol{x})\,\rme^{-iH_{\scriptscriptstyle Q} (t-t_0)}$, with $n^{\scriptscriptstyle A}(\x)$ given by Eq.~(\ref{colourdensity}).  Finally, the QGP  correlators, which result from the integration over the plasma degrees of freedom, are defined as:
	\begin{align}
		g^{2}\left\langle A_{0}^{\scriptscriptstyle B}\left(t,\boldsymbol{x}\right)A_{0}^{\scriptscriptstyle C}\left(t',\boldsymbol{x}^{\prime}\right)\right\rangle _{0}&=\delta^{\scriptscriptstyle BC}\Delta^{>}\left(t-t',\boldsymbol{x}-\boldsymbol{x}^{\prime}\right)\nonumber\\
		g^{2}\left\langle A_{0}^{\scriptscriptstyle C}\left(t',\boldsymbol{x'}\right)A_{0}^{\scriptscriptstyle B}\left(t,\boldsymbol{x}\right)\right\rangle _{0}&=\delta^{\scriptscriptstyle BC}\Delta^{<}\left(t-t',\boldsymbol{x}-\boldsymbol{x}^{\prime}\right)
		\label{thermal-propagators}
	\end{align} 
	where $\langle \cdots\rangle_0$ denotes the average with the QGP equilibrium density matrix \cite{blaizot2018quantum}. 
	These correlators satisfy the following relations, easily verified by inspection, 
	\begin{eqnarray}\label{eq:relationsDelta}
		\Delta^{<}\left(t-t',\boldsymbol{x}-\boldsymbol{x}'\right)&=\Delta^{>}\left(t'-t,\boldsymbol{x}'-\boldsymbol{x}\right)=\Delta^{>\star}\left(t-t',\boldsymbol{x}-\boldsymbol{x}'\right). 
	\end{eqnarray}
	
	The derivation of Eq.~(\ref{eq:be}) assumes that the interaction between the plasma and the heavy quarks is weak, justifying an expansion up to (formally) second order in $H_1$ as well as the  factorised form of the full  density matrix into a tensor product of the density matrix of the heavy quarks and that of the plasma. To within additional  adjustments  that will be implemented as we proceed (noticeably the substitution of $\bar \rho(t')$ by $\bar \rho(t)$ in its r.h.s.), Eq.~(\ref{eq:be}) is essentially the Redfield equation, and the underlying approximation is commonly referred to as the Born-Markov approximation (see e.g. \cite{breuer2002theory}). 
	
	As shown in \cite{nathan2020universal}, simple manipulations, compatible with the Born-Markov approximation, allow us to rewrite the master equation (\ref{eq:be}) in  such a way  that it acquires the form of a Lindblad equation. To do so, 
	we first write the plasma correlator (\ref{thermal-propagators}) as a convolution of so-called ``jump correlators"  $g\left(t,\boldsymbol{x}\right)$:
	\begin{equation}
		\Delta^{>}\left(t-t^{\prime},\boldsymbol{x}-\boldsymbol{x}^{\prime}\right)=\int_{-\infty}^{+\infty}\text{d}v\int_{\boldsymbol{y}}g\left(t-v,\boldsymbol{x-y}\right)\,g\left(v-t^{\prime},\boldsymbol{y}-\boldsymbol{x}^{\prime}\right),\label{convolution-ULEa}
	\end{equation}
	where we assume  that $g(-t,-\boldsymbol{x})=g^*(t,\boldsymbol{x})$, a sufficient condition for $\Delta^{>}(t,\x)$ to satisfy the relations (\ref{eq:relationsDelta}). This property will be used repeatedly in the following. It is useful to note that $(\x,t)$ and $(\x',t')$ are the space-time coordinates of the points where a gluon is hooked, while $(\y, v)$ are the space-time coordinates of the intermediate point introduced in the convolution (\ref{convolution-ULEa}). 
	
	In Fourier space, the relation (\ref{convolution-ULEa}) has a simple expression
	\begin{equation}
		\Delta^{>}\left(q_{0},\boldsymbol{q}\right)=\left[g\left(q_{0},\boldsymbol{q}\right)\right]^{2}\label{eq:10},
	\end{equation}
	where we have defined\footnote{With a slight abuse of notation, we use the same notation for the (jump) correlator and its Fourier transform.}
	\begin{equation}
		g\left(t,\boldsymbol{x}\right)=\int\frac{\rmd q_0}{2\pi}\int_{\q}\rme^{-iq_{0} t+i \boldsymbol{q}\cdot\boldsymbol{x}} g(q_0,\boldsymbol{q}), \label{g-as-function-of-Deltaa}
	\end{equation}
	Note that the condition $g(-t,-\boldsymbol{x})=g^*(t,\boldsymbol{x})$ implies that $g(q_0,\q)$ is real.
	
	At this point we recall that the response of the plasma to perturbations occurs on a time scale $\tau_E$  which is essentially the inverse of the Debye mass $m_D$. This is supposed to be small compared to the typical scale $\tau_R$ that characterises the relaxation dynamics of the $Q\bar{Q}$ reduced density matrix $\bar{\rho}$ when this pair is in contact with the QGP,\footnote{See e.g. \cite{akamatsu2022quarkonium,Delorme:2024rdo} for qualitative and quantitative estimates.} i.e. $\tau_E\ll \tau_R$. Furthermore, we are interested in the long-time dynamics, that is we want to follow the evolution of the system over time scales that are at least of order $\tau_R$, ignoring possible transients near $t_0$. Let us then integrate Eq.~(\ref{eq:be}) from    $t_0$ to $t_1$, with $|t_1-t_0|\gtrsim\tau_R\gg \tau_E$, and substitute Eq.~(\ref{convolution-ULEa}). We obtain  
	\begin{equation}
		\bar\rho\left(t_1\right)-\bar\rho\left(t_0\right) =-  \int_{t_0}^{t_1}\rmd t \int_{-\infty}^{+\infty} \rmd v \int_{t_0}^{t}\rmd t'\,g\left(t-v\right)g\left(v-t^{\prime}\right)
		\left[\bar n^{\scriptscriptstyle A}\left(t\right),\bar n^{\scriptscriptstyle A}\left(t^{\prime}\right)\bar \rho\left(t^{\prime}\right)\right]+\mathrm{h.c.}, 
		\label{eq:intbea}
	\end{equation}
	where we have exchanged the order of the integrations over $v$ and $t'$.  Note that since the spatial coordinates play no role in this and the forthcoming manipulations, we omit them temporarily in order to alleviate the notation. They will be reinstated in the final expressions.
	
	\begin{figure}
		\centering\includegraphics[width=0.45\textwidth]{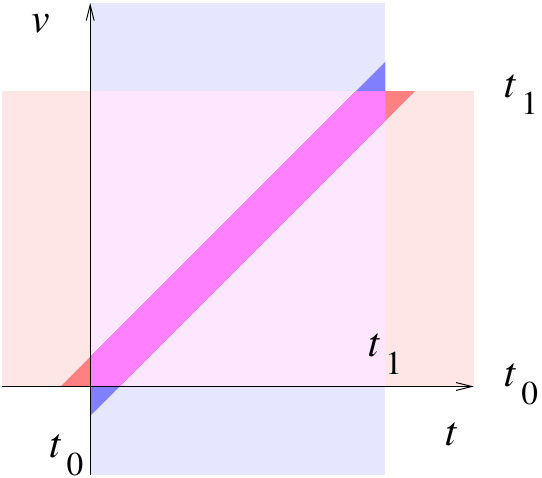}
		\caption{Domains of integration corresponding to Eqs.~(\ref{eq:intbea}) [blue] and~(\ref{eq:intbe2a}) [red]. The band corresponding to $|t-v|<\tau_E$ is shown with darker colours.}
		\label{fig:domaintvb}
	\end{figure}

	We note now that since $g(t-v)$ has support only in the strip $|t-v|\lesssim\tau_E$ of the $(t,v)$ plane,  integrating over $v$ at fixed $t$ yields the same result as integrating over $t$ at fixed $v$, to within contributions at the boundaries of the integration domain that are  negligible if $|t_1-t_0|\gg \tau_E$ (see Fig.~\ref{fig:domaintvb}). In other words, and this is the crucial step in the argument,  we can exchange the domains of integration over $v$ and $t$, namely perform the substitution $\int_{t_0}^{t_1}\rmd t\int_{-\infty}^\infty \rmd v\mapsto \int_{t_0}^{t_1}\rmd v\int_{-\infty}^\infty \rmd t$, and  rewrite Eq.~(\ref{eq:intbea}) as
	\begin{equation}
		\bar \rho\left(t_1\right)-\bar \rho\left(t_0\right) =-  \int_{t_0}^{t_1}\rmd v \int_{-\infty}^{+\infty}\text{d}t \int_{t_0}^{t}\rmd t'\,g\left(t-v\right)g\left(v-t^{\prime}\right)
		\left[\bar n^{\scriptscriptstyle A}\left(t\right),\bar n^{\scriptscriptstyle A}\left(t^{\prime}\right) \bar\rho\left(t^{\prime}\right)\right]+\mathrm{h.c.} + \cdots, 
		\label{eq:intbe2a}
	\end{equation}
	where ``$\cdots$" represent small corrections, as the same order of those neglected while performing the Born-Markov approximation leading to the Redfield equation; see~\cite{nathan2020universal} for more details. Neglecting these corrections, we take, as a next step,  the derivative with respect to $t_1$ and get
	\begin{equation}
		\frac{\rmd\bar\rho \left(t_1\right)}{\rmd t_1} =- \int_{-\infty}^{+\infty}\text{d}t   \int_{t_{0}}^{t}dt^{\prime} g\left(t-t_1\right)g\left(t_1-t^{\prime}\right)\left[\bar n^{\scriptscriptstyle A}\left(t\right),\bar n^{\scriptscriptstyle A}\left(t^{\prime}\right)\bar\rho\left(t_1\right)\right]+\text{h.c.}, 
		\label{eq:2.17}
	\end{equation} 
	where, consistently with the Born-Markov approximation, we have replaced $\bar \rho(t')$ by $\bar \rho(t_1)$ in the right-hand side of the equation ($|t_1-t'|\lesssim \tau_E$, and $\bar\rho(t_1+\tau_E)\simeq \bar\rho(t_1)$). This renders the equation local in $t_1$.   Finally, by reinstating the spatial coordinates, and relabelling the time integration variables, we obtain
	\begin{equation}
		\begin{split}
			\frac{\rmd\bar \rho\left(t\right)}{\rmd t}=&-\int_{-\infty}^{+\infty}\text{d}v\int_{t_0}^{+\infty}\text{d}v^{\prime}\,\theta(v-v')\int_{\x\x'\y}g(v-t,\x-\y)g\left(t-v^{\prime},\boldsymbol{y}-\boldsymbol{x}^{\prime}\right)\\[2ex]
			&\times \left[\bar n^{\scriptscriptstyle A}\left(v,\boldsymbol{x}\right),\bar n^{\scriptscriptstyle A}(v',\x')\bar\rho\left(t\right)\right]+\text{h.c.}
		\end{split}
		\label{eq:2.18}
	\end{equation}
	where we have used the theta function to extend the upper bound of the $v'$ integration to $+\infty$. Note that, while the previous manipulations have destroyed the initial convolution structure of the time integration in the definition (\ref{convolution-ULEa}), this convolution remains intact for the spatial variables.  The purpose of this manipulation is to allow us to assemble a factor $g$ and a density $n^{\scriptscriptstyle A}$ into a Lindblad operator, as we shall now show. 
	
	At this point it is convenient to return to the Schrödinger representation. Recalling that $\rho(t)=U_{\scriptscriptstyle Q}(t-t_0) \bar \rho(t) U^\dagger_{\scriptscriptstyle Q}(t-t_0)$ with $U_{\scriptscriptstyle Q}(t-t_0)=e^{-i H_{\scriptscriptstyle Q}(t-t_0)}$,
	we get
	\beq\label{eq:cahngefromIto S}
	\frac{\rmd \rho}{\rmd t}=-i[H_{\scriptscriptstyle Q},\rho]+U_{\scriptscriptstyle Q}(t-t_0) \frac{\rmd \bar\rho(t)}{\rmd t} U^\dagger_{\scriptscriptstyle Q}(t-t_0),
	\eeq
	and similarly, for the colour densities,
	\beq
	U_{\scriptscriptstyle Q}(t-t_0) \bar n^{\scriptscriptstyle A}(v,\x) U^\dagger_{\scriptscriptstyle Q}(t-t_0)&=& U_{\scriptscriptstyle Q}(t-t_0)U^\dagger_{\scriptscriptstyle Q}(v-t_0)n^{\scriptscriptstyle A}(\x) U_{\scriptscriptstyle Q}(v-t_0) U^\dagger_{\scriptscriptstyle Q}(t-t_0)  \nonumber\\ 
	&=& U_{\scriptscriptstyle Q}^\dagger(v-t) n^{\scriptscriptstyle A}(\x) U_{\scriptscriptstyle Q}(v-t).
	\eeq
	After this transformation, any explicit dependence on the initial time $t_0$ has disappeared. Since we are not interested in transient phenomena, we can push this initial time to the remote past, i.e. let $t_0\to -\infty$ in the integral over $v'$. A simple calculation then yields 
	\beq
	\frac{\rmd \rho}{\rmd t}+ i[H_{\scriptscriptstyle Q},\rho]=-\int_{v v'}\,\theta\left(v-v'\right)\int_{\y} \left[\ell^{{\scriptscriptstyle A}\,\dagger}(t-v,\y), \ell^{\scriptscriptstyle A}(t-v',\y)\rho(t)\right]+\text{h.c.}
	\label{eq:2.18ba}
	\eeq
	where we have used the shorthand $\int_{-\infty}^\infty \rmd v\mapsto\int_v$ and set 
	\beq
	\ell^{\scriptscriptstyle A}(t-v,\y)\equiv\int_\x g(t-v,\boldsymbol{y-x}) U_{\scriptscriptstyle Q}^\dagger(v-t) n^{\scriptscriptstyle A}(\x) U_{\scriptscriptstyle Q}(v-t).
	\eeq
	We then define the following Lindblad operators
	\begin{equation}
		L^{\scriptscriptstyle A}(\y) =\int_{-\infty}^{+\infty}\rmd v\, \int_\x g(t-v,\boldsymbol{y-x}) U_{\scriptscriptstyle Q}^\dagger(v-t) n^{\scriptscriptstyle A}(\x) U_{\scriptscriptstyle Q}(v-t).\label{Lindblad-operator-interactionA}
	\end{equation}
	The hermitian conjugate $ L^{{\scriptscriptstyle A}\dagger}(\y)$ differs from $ L^{\scriptscriptstyle A}(\y)$ solely by the substitution of $g(t-v,\boldsymbol{y-x})$ by $g^*(t-v,\boldsymbol{y-x})$.
	
	To proceed further, it is convenient to write the theta function in Eq.~(\ref{eq:2.18ba}) as 
	\begin{equation}
		\theta\left(v-v^{\prime}\right)=\frac{1}{2}+\frac{1}{2}\,\text{sign}\left(v-v^{\prime}\right).
	\end{equation}
	The first term, 1/2, yields 
	\beq\label{eq:Lindblad0}
	-\frac{1}{2}\int_{\y}\left[L^{{\scriptscriptstyle A}\,\dagger}(\y),L^{\scriptscriptstyle A}(\y)\rho\left(t\right)\right]+\text{h.c.}
	=\int_\y L^{\scriptscriptstyle A}(\y)\rho(t) L^{{\scriptscriptstyle A}\,\dagger}(\y))-\frac{1}{2}\left\{ L^{{\scriptscriptstyle A}\,\dagger}(\y))L^{\scriptscriptstyle A}(t,\y) ,\rho(t)\right\}.\nonumber\\
	\eeq
	This is the part of the equation that contributes to the non-unitary evolution. 
	
	The term which involves the sign function can be rewritten as follows
	\beq
	&&\int_{v v'}\text{sign}\left(v-v^{\prime}\right)\int_{\y} \left[\ell^{{\scriptscriptstyle A}\,\dagger}(t-v,\y), \ell^{\scriptscriptstyle A}(t-v',\y)\rho\left(t\right)\right]+\text{h.c.}\nonumber\\
	&& =\int_{v v'}\text{sign} (v-v')\,\int_{\y} \left[\ell^{{\scriptscriptstyle A}\,\dagger}(t-v,\y) \ell^{\scriptscriptstyle A}(t-v',\y),\rho(t)\right]
	\eeq
	In order to simplify this expression we use the following representation of the sign function
	\begin{equation}
		\frac{1}{2}{\rm sign}(v-v')=\frac{i}{2\pi } P\int_{-\infty}^{+\infty} \frac{d\omega}{\omega} \rme^{-i \omega[(v-t)-(v'-t))},
	\end{equation} 
	where $P\int$ denotes the principal value of the integral, and we define 
	\begin{equation}
		L^{\scriptscriptstyle A}(\omega,\y) = \int_{-\infty}^{+\infty} \rmd v\, \rme^{i \omega(v-t)} l^{\scriptscriptstyle A}(t-v,\y),\qquad L^{\scriptscriptstyle A}(\y)=L^{\scriptscriptstyle A}(\omega=0,\y). 
	\end{equation} 
	We then obtain
	\beq
	&&-\frac{1}{2}  \int_{v v'}\text{sign} (v-v')\,\int_{\y}\left[\ell^{{\scriptscriptstyle A}\,\dagger}(t-v,\y) \ell^{\scriptscriptstyle A}(t-v',\y),\rho(t)\right]\nonumber\\
	&&=-\frac{i}{2\pi} P\int\frac{\rmd \omega}{\omega}\int_\y [L^{{\scriptscriptstyle A}\,\dagger}(\omega,y) L^{\scriptscriptstyle A}(\omega,y),\rho(t)].
	\eeq
	This term is hermitian and plays the role of a correction to the hermitian hamiltonian $H_{\scriptscriptstyle Q}$, $H_{\scriptscriptstyle Q}\mapsto H_{\scriptscriptstyle Q}+\Lambda$ where 
	\begin{equation}\label{eq:Lambda}
		\Lambda = \frac{1}{2\pi}P \int \frac{d\omega}{\omega} \int_y L^{{\scriptscriptstyle A}\,\dagger}(\omega,y) L^{\scriptscriptstyle A}(\omega,y).
	\end{equation}
	
	In summary, the QCD Nathan-Rudner ULE takes the form of the following Lindblad equation
	\begin{equation}
		\frac{d\rho\left(t\right)}{dt}=-i\left[H_{\scriptscriptstyle Q}+\Lambda,\rho\left(t\right)\right]+\int_{\boldsymbol{y}}\left(L^{{\scriptscriptstyle A}}\left(\boldsymbol{y}\right)\rho\left(t\right)L^{{\scriptscriptstyle A}\,\dagger}\left(\boldsymbol{y}\right)-\frac{1}{2}\left\{ L^{{\scriptscriptstyle A}\,\dagger}\left(\boldsymbol{y}\right)L^{{\scriptscriptstyle A}}\left(\boldsymbol{y}\right),\rho\left(t\right)\right\} \right)
		\label{eq:ULEsch}
	\end{equation}
	where the Lindblad operator  is defined in (\ref{Lindblad-operator-interactionA}), and the correction to the hamiltonian given in (\ref{eq:Lambda}).  
	
	\subsection{The Lindblad operators and their colour structure}
	
	The state of a heavy quark can be characterised by a position, a colour, and a spin. We ignore here  the spin degree of freedom. Then the reduced density matrix $\rho$ has matrix elements of the form
	\beq
	\langle \r_1 a, \bar\r_1 \bar a |\rho|\r_2 b,\bar\r_2 \bar b\rangle,
	\eeq
	where $a, b$  and $\bar a, \bar b$   are colour indices in the fundamental representation and  its conjugate, respectively, while $\r_i$ and $\bar\r_i$  ($i=1,2$) denote respectively  the coordinates of the quark and the antiquark. 
	Factorising the colour structure, one can write $\rho$ as follows (see \cite{blaizot2018quantum,blaizot2018approach} for more details on the colour structure of $\rho$): 
	\begin{equation}
		\rho(t)=\left(\frac{\delta_{a\bar{a}}\delta_{b\bar{b}}}{N_c}\rho_{\rm s}(t)+\frac{t^{\scriptscriptstyle\it A}_{a\bar{a}}t^{\scriptscriptstyle\it A}_{\bar{b}b}}{T_F}\rho_{\rm o}(t)\right)|a,\bar{a}\rangle\langle b,\bar{b}|
		= \rho_{\rm s}(t) \ket{\rm s}\bra{\rm s}+\rho_{\rm o}(t) \sum_{\scriptscriptstyle C} \ket{\rm o^{\scriptscriptstyle C}}\bra{\rm o^{\scriptscriptstyle C}}
		\label{eq:rhodecom1}
	\end{equation}
	where $\rho_{\rm s }$ and $\rho_{\rm o}$ are matrices in the $Q\bar Q$ coordinate space, i.e. the matrix elements of $\rho_{\rm s}$ are $\langle \r_1, \bar\r_1 |\rho_{\rm s}|\r_2 ,\bar\r_2 \rangle$, and similarly for $\rho_{\rm o}$. In the formula above, $T_F=1/2$ and the colour matrices are normalised as ${\rm Tr}\,t^{\scriptscriptstyle A} t^{\scriptscriptstyle B}=\delta^{\scriptscriptstyle AB}/2$. The relation between the second and third members of Eq.~(\ref{eq:rhodecom1}) results from the following formulae 
	\beq
	\bra{a\bar a}{\rm s}\rangle=\delta_{a\bar a}\frac{1}{\sqrt{N_c}},\qquad  \bra{a\bar a}{\rm o^{\scriptscriptstyle C}}\rangle=\sqrt{2} \,t^{\scriptscriptstyle C}_{a\bar a},
	\label{eq:colour-identities}
	\eeq
	where $\ket{\rm s}$ and $\ket{{\rm o}^{\scriptscriptstyle C}}$ denote respectively colour singlet and octet (normalised) states, the index ${\scriptstyle\it C}$ in  ${\rm o}^{\scriptscriptstyle\it C}$  distinguishing the various members of the octet representation. Note that $\rho_{\rm o}$ does not depend on the member of the octet, and we have ${\rm Tr}\rho={\rm Tr}\rho_{\rm s}+(N_c^2-1){\rm Tr}\rho_{\rm o}$, where here the trace does not include the colour degrees of freedom.
	
	Similarly the Lindblad operator (\ref{Lindblad-operator-interactionA})
	is an operator in the $Q\bar Q$ Hilbert space, with matrix elements of the form 
	\beq\label{Lindblad-operator-mel}
	&&\langle \r_1a, \bar\r_1 \bar a |L^{\scriptscriptstyle A}(\y)|\r_2 b,\bar\r_2 \bar b\rangle   =\int_{-\infty}^{+\infty}\rmd v\, \int_\x g(t-v,\boldsymbol{y-x})\nn 
	&&\qquad\qquad\qquad\qquad\times\langle \r_1a, \bar\r_1 \bar a |U^\dagger(v-t) n^{\scriptscriptstyle A}(\x) U(v-t))|\r_2 b,\bar\r_2 \bar b\rangle, 
	\eeq 
	where we omit, from now on, the subscript $Q$ on the evolution operator to alleviate the notation. In fact, this operator is diagonal in the singlet-octet basis~\cite{blaizot2018approach} and we will add a subscript to denote the corresponding diagonal elements,  respectively  $U_{\text{s}}(v)=\rme^{-iv H_{\text{s}}}$ and $U_{\text{o}}(v)=\rme^{-iv H_{\text{o}}}$,  with $H_{\text{s}}$ and $H_{\text{o}}$ the hamiltonians of the heavy quark-antiquark pair in the corresponding singlet and octet channels.
	
	Since the evolution  operators are diagonal in colour, the  colour structure of $L^{\scriptscriptstyle A}$ is determined by that of the density operator $n^{\scriptscriptstyle A}$, whose expression is given in Eq.~(\ref{colourdensity}). 
	Its matrix elements are written as 
	\beq\label{eq:matrixelem1}
	\bra{{\rm s}} n^{\scriptscriptstyle A}(\x)\ket{{\rm o}^{\scriptscriptstyle C}}=\bra{{\rm o}^{\scriptscriptstyle C}} n^{\scriptscriptstyle A}(\x)\ket{{\rm s}}=\frac{\delta^{{\scriptscriptstyle AC}} }{\sqrt{2N_c}}\,  n(\x).
	\eeq
	and 
	\beq\label{eq:matrixelem2}
	\bra{{\rm o}^{\scriptscriptstyle D}}n^{\scriptscriptstyle A}(\x)\ket{{\rm o}^{\scriptscriptstyle C}}=\frac{1}{2} d^{\scriptscriptstyle DAC}\,n(\x)+\frac{i}{2}f^{\scriptscriptstyle DAC}  \,m(\x),
	\eeq
	where  
	\beq
	n(\x)&\equiv &\delta(\x-\hat\r)\otimes\mathbb{I} -\mathbb{I}\otimes \delta(\x-\hat\r),\nn  m(\x)&\equiv&\delta(\x-\hat\r)\otimes\mathbb{I} +\mathbb{I}\otimes \delta(\x-\hat\r).
	\eeq
	The operators $n(\x)$ and $m(\x)$ are local operators that act on the spatial coordinates. We have 
	\beq
	\bra{\r_1,\r_2} n(\x)\ket{\r_3,\r_4}=
	\delta(\r_1-\r_3)\delta(\r_2-\r_4) \left[ \delta(\x-\r_1)-\delta(\x-\r_2)\right],
	\eeq
	and similarly for $m(\x)$. Since the evolution operators in the Lindblad operator (\ref{Lindblad-operator-mel}) change the values of the spatial coordinates, it is convenient to introduce next to the density operator a projector that specifies the values of these coordinates on which $n(\x)$ or $m(\x)$ are acting. Thus we define   $P_{\boldsymbol{X}}\equiv\left|\boldsymbol{X}\right\rangle \left\langle \boldsymbol{X}\right|\equiv\left|\boldsymbol{x},\bar{\boldsymbol{x}}\right\rangle \left\langle \boldsymbol{x},\bar{\boldsymbol{x}}\right|$, where $\left(\boldsymbol{x},\bar{\boldsymbol{x}}\right)$ refers to the quark and antiquark positions, respectively. The effect  of this projector is to specify the value of $\x$ in the function $g(t-v,\boldsymbol{y-x})$. In case $n(\x)$, respectively $m(\x)$, is sitting next to the projector, the function $g(t-v,\boldsymbol{y-x})$ is then substituted respectively  by $g_-(t-v,\boldsymbol{y-x})\equiv g(t-v,\boldsymbol{y-x})-g(t-v,\boldsymbol{y-\bar x})$, or by $g_+(t-v,\boldsymbol{y-x})\equiv g(t-v,\boldsymbol{y-x})+g(t-v,\boldsymbol{y-\bar x})$). 
	
	Collecting all the previous results, we then obtain the following expressions for the relevant matrix elements in colour space of the Lindblad operators:
	\begin{align}
		\bra{{\rm s}} L^{\scriptscriptstyle A}(\y)\ket{ {\rm o}^{\scriptscriptstyle C} } &= \frac{\delta^{\scriptscriptstyle AC} }{  \sqrt{2N_{c}} } \int_{v\boldsymbol{X}} g_-\left(t-v,\boldsymbol{y-X}\right) U^\dagger_{\rm s}\left(v-t\right)P_{\boldsymbol{X}}U_{\text{o}}\left(v-t\right),
		\label{eq:L0}\\
		\bra{{\rm o^{\scriptscriptstyle C}}} L^{\scriptscriptstyle A}(\y)\ket{ \rm s}  &= \frac{\delta^{\scriptscriptstyle AC} }{  \sqrt{2N_{c}} } \int_{v\boldsymbol{X}} g_-\left(t-v,\boldsymbol{y-X}\right)
		U^\dagger_{\rm o}\left(v-t\right)P_{\boldsymbol{X}}U_{\text{s}}\left(v-t\right),
		\label{eq:L1}\\
		\bra{{\rm o}^{\scriptscriptstyle D}} L^{\scriptscriptstyle A}(\y)\ket{ {\rm o}^{\scriptscriptstyle C} } &= \frac{1}{2} \int_{v\boldsymbol{X}}\bigl[d^{\scriptscriptstyle DAC}   g_-(t-v,\boldsymbol{y}-\boldsymbol{X}) +i f^{\scriptscriptstyle DAC} g_+(t-v,\boldsymbol{y}-\boldsymbol{X})\bigr]\nn
		&\qquad\qquad\qquad\qquad\qquad\qquad\qquad\times 
		U^\dagger_{\rm o}\left(v-t\right)P_{\boldsymbol{X}}U_{\rm o}\left(v-t\right).
		\label{eq:L2}
	\end{align}
	The matrix  element $\bra{{\rm s}} L^{\scriptscriptstyle A}(\y)\ket{ {\rm o}^{\scriptscriptstyle C} } $ can be read as follows. A $Q\bar Q$ pair in an octet state propagates as an octet from time $t$ to time $v$  where the density operator is acting. This operator changes the colour state of the pair from octet to singlet (a gluon is emitted or absorbed). The singlet state then propagates backwards from $v$ to $t$. The same interpretation holds for the octet to octet transitions, with the additional feature that the symmetric and antisymmetric states contribute with their appropriate colour factors. 
	
	\subsection{The reduced Lindblad operators} 
	
	In order to perform the colour algebra involved in the summation over $A$ in the Lindblad equation (\ref{eq:ULEsch}), which will eventually lead to reduced Lindbad operators, we note that both the density matrix and the heavy quark hamiltonian (including its $\Lambda$ contribution)\footnote{Quite generally, the combination $\sum_{\scriptscriptstyle A} t^{\scriptscriptstyle A} t^{\scriptscriptstyle A}$ contained in $\Lambda$ does not induce colour transitions.} are diagonal in the singlet-octet basis, while the Lindblad equation couples $\rho_{\rm s}$ and $\rho_{\rm o}$. The overall colour structure is best visualised by writing the coupled equations for $\rho_{\rm s}$ and $\rho_{\rm o}$ as follows
	\beq \label{eq:singlet-equation-app}
	&&\frac{d\rho_{\rm s}\left(t\right)}{dt}=-i\left[H_{\rm s}+\Lambda_{\rm s}\left(t\right),\rho_{\rm s}\left(t\right)\right]+\mathcal{L}^{\rm ss}\left(t\right)\rho_{\rm s}\left(t\right)+\mathcal{L}^{\rm so}\left(t\right)\rho_{\rm o}\left(t\right)\\
	&&\frac{\rmd\rho_{\rm o}(t)}{\rmd t}=-i\left[H_{\rm o}+\Lambda_{\rm o}\left(t\right),\rho_{\rm o}\left(t\right)\right]+\mathcal{L}^{\rm os}\left(t\right)\rho_{\rm s}\left(t\right)+\mathcal{L}^{\rm oo}\left(t\right)\rho_{{\rm o}}\left(t\right).
	\eeq
	where the operators $\mathcal{L}^{\rm ss}$, etc, can be viewed as Liouvilian operators in all quantum numbers of the two particle states, colour aside. In the following, we shall obtain expressions for these transitions operators in terms of reduced Lindblad operators. 
	
	To proceed, we note first that the density operator $n^{\scriptscriptstyle A}(\x)$ connects singlet to octet states, and also various octet states among themselves, with matrix elements given in Eqs.~(\ref{eq:matrixelem1}) and (\ref{eq:matrixelem2}). 
	The calculation then proceeds easily by inserting closure relations in the singlet-octet basis at appropriate places and using the relations
	\beq
	&&\bra{{\rm s}}n^{\scriptscriptstyle A}(\x) n^{\scriptscriptstyle A}(\x')\ket{{\rm s}}=\sum_C\bra{{\rm s}}n^{\scriptscriptstyle A}(\x)\ket{{\rm o}^{\scriptscriptstyle C}}\bra{{\rm o}^{\scriptscriptstyle C}} n^{\scriptscriptstyle A}(\x')\ket{{\rm s}},\nn
	&&\bra{{\rm o}^{\scriptscriptstyle C}}n^{\scriptscriptstyle A}(\x) n^{\scriptscriptstyle A}(\x')\ket{{\rm o}^{\scriptscriptstyle C}}=\bra{{\rm o}^{\scriptscriptstyle C}} n^{\scriptscriptstyle A}(\x)\ket{{\rm s}}\bra{{\rm s}}n^{\scriptscriptstyle A}(\x')\ket{{\rm o}^{\scriptscriptstyle C}}+\sum_{\scriptscriptstyle D}\bra{{\rm o}^{\scriptscriptstyle C}}n^{\scriptscriptstyle A}(\x)\ket{{\rm o}^{\scriptscriptstyle D}}\bra{{\rm o}^{\scriptscriptstyle D}} n^{\scriptscriptstyle A}(\x')\ket{{\rm o}^{\scriptscriptstyle C}},
	\nn
	\eeq
	as well as the  following formulae
	\beq\label{colalgebra}
	f^{\scriptscriptstyle ABC} f^{\scriptscriptstyle ABD}=N_c\,\delta^{\scriptscriptstyle CD}, \qquad d^{\scriptscriptstyle ABC}d^{\scriptscriptstyle ABD}=\frac{N_c^2-4}{N_c}\delta^{\scriptscriptstyle CD}, \qquad d^{\scriptscriptstyle ABC}f^{\scriptscriptstyle ABD}=0.
	\eeq
	Consider then the equation for $\rho_{\rm s}$, and focus first on the term   $\left\{ L^{{\scriptscriptstyle A}\,\dagger}(\y)L^{\scriptscriptstyle A}(\y) ,\rho(t)\right\}$. By taking a matrix element between singlet states we get, for the first term in the anticommutator, 
	\beq
	\bra{{\rm s}}L^{{\scriptscriptstyle A}\dagger}(\y)L^{\scriptscriptstyle A}(\y) \rho_{\rm s}\ket{\rm s}&=&\sum_{\scriptscriptstyle C}\bra{{\rm s}}L^{{\scriptscriptstyle A}\dagger}(\y)\ket{{\rm o}^{\scriptscriptstyle C}}\bra{{\rm o}^{\scriptscriptstyle C}} L^{\scriptscriptstyle A}(\y)\rho_{\rm s}\ket{{\rm s}}\nn
	&=& C_F \int_{v\boldsymbol{X},v'\boldsymbol{X'} } g_-^*\left(t-v,\boldsymbol{y-X}\right)g_-\left(t-v',\boldsymbol{y-X'}\right)\nn
	&\times& U^\dagger_{\rm s}\left(v-t\right)P_{\boldsymbol{X}}U_{\rm o}(v-t)U_{\rm o}^\dagger(v'-t)P_{\boldsymbol{X'}}U_{\text{s}}\left(v'-t\right)\rho_{\rm s},
	\eeq 
	where the summation over the repeated colour index $A$ has been used. Note that the last line remains an operator in the $Q\bar Q$ coordinate space. The factorised structure of the integrand allows us to identify a reduced Lindblad operator $L_{[\text{os}]}(\y)$ which we define as 
	\beq\label{eq:Los}
	L_{[\text{os}]}(\y)\equiv \int_{v',X'}g_-\left(t-v',\boldsymbol{y-X'}\right) U^\dagger_{\rm o}\left(v'-t)\right)P_{\boldsymbol{X'}}U_{\text{s}}\left(v'-t\right),
	\eeq
	so that 
	\beq
	\bra{{\rm s}}L^{{\scriptscriptstyle A}\dagger}(\y)L^{\scriptscriptstyle A}(\y) \rho_{\rm s}\ket{{\rm s}}=C_F L^\dagger_{[\text{os}]} L_{[\text{os}]}\rho_{\rm s}.
	\eeq
	The notation $L_{[\text{os}]}$ is meant to recall that this reduced operator emerges from the matrix element $\bra{{\rm o^{\scriptscriptstyle C}}} L^{\scriptscriptstyle A}(\y)\ket{ \rm s}$ in Eq.~(\ref{eq:L1}) that connects the two irreducible colour representations, but $L_{[\text{os}]}$ does not act on colour states anymore. That is, $L_{[\text{os}]}$ is not an operator in colour space.
	A similar analysis holds for the other member of the anticommutator. This  allows us to write the corresponding contribution to ${\cal L}^{\rm ss}\rho_{\rm s} $ as follows
	\beq
	-\frac{1}{2}\bra{{\rm s}}L^{{\scriptscriptstyle A}\dagger}(\y)L^{\scriptscriptstyle A}(\y) \rho_{\rm s}\ket{{\rm s}}=-\frac{C_F}{2}\left\{ L^\dagger_{[\text{os}]} L_{[\text{os}]},\rho_{\rm s}\right\}.
	\eeq
	Consider next the following term,
	\beq
	\bra{{\rm s}}L^{{\scriptscriptstyle A}}(\y)\rho(t)L^{{\scriptscriptstyle A \dagger}}(\y) \ket{{\rm s}}&=&\sum_{\scriptscriptstyle C }\bra{{\rm s}}L^{{\scriptscriptstyle A}}(\y)\ket{{\rm o}^{\scriptscriptstyle C}} \rho_{\rm o}(t)  \bra{{\rm o}^{\scriptscriptstyle C}} L^{\scriptscriptstyle A \dagger}(\y)\ket{{\rm s}}\nn
	&=& C_F \int_{v\boldsymbol{X},v'\boldsymbol{X'} } g_-\left(t-v,\boldsymbol{y-X}\right)g_-^*\left(t-v',\boldsymbol{y-X'}\right)\nn
	&\times& U^\dagger_{\rm s}\left(v-t\right)P_{\boldsymbol{X}}U_{\rm o}(v-t) \rho_{\rm o}(t) U^\dagger_{\rm o} (v'-t)P_{\boldsymbol{X'}}U_{\text{s}}\left(v'-t\right)\nn
	&=& C_F L_{[\rm so]} \rho_{\text{o}} L^\dagger_{[\rm so]},
	\eeq
	with 
	\beq\label{eq:Lso}
	L_{[\rm so]}(\y)\equiv \int_{v,\boldsymbol{X}} g_-\left(t-v,\boldsymbol{y-X}\right)U^\dagger_{\rm s}\left(v-t\right)P_{\boldsymbol{X}}U_{\rm o}(v-t).
	\eeq 
	This term couples $\rho_{\rm s}$ and $\rho_{\rm o}$, and contributes therefore to ${\cal L}^{\rm so}\rho_{\rm o} $.
	
	Consider next the evolution of the octet density
	$\dot{\rho}_{\rm o}:=  
	\bra{{\rm o}^{\scriptscriptstyle C}}\dot{\rho}(t) \ket{{\rm o}^{\scriptscriptstyle C}}$ --- where $\ket{{\rm o}^{\scriptscriptstyle C}}$ is any element of the octet representation --- and focus first on the term $\bra{{\rm o}^{\scriptscriptstyle C}}L^{{\scriptscriptstyle A}}(\y)\rho(t)L^{{\scriptscriptstyle A \dagger}}(\y) \ket{{\rm o}^{\scriptscriptstyle C}}$ 
	associated to Eq.~(\ref{eq:ULEsch}). Inserting the identity (in colour space) written as the sum of singlet and octet projectors, one obtains
	\begin{equation}
		\bra{{\rm o}^{\scriptscriptstyle C}}
		L^{{\scriptscriptstyle A}}(\y)\rho L^{\scriptscriptstyle A \dagger}(\y) \ket{{\rm o}^{\scriptscriptstyle C}}=\bra{{\rm o}^{\scriptscriptstyle C}}L^{\scriptscriptstyle A}(\y)\ket{s}\rho_{\rm s}\bra{s}L^{\scriptscriptstyle A \dagger}(\y) \ket{{\rm o}^{\scriptscriptstyle C}}
		+ \sum_{\scriptscriptstyle D}
		\bra{{\rm o}^{\scriptscriptstyle C}}L^{\scriptscriptstyle A}(\y)
		\ket{{\rm o}^{\scriptscriptstyle D}}\rho_{\rm o}\bra{{\rm o}^{\scriptscriptstyle D}}L^{\scriptscriptstyle A \dagger}(\y) \ket{{\rm o}^{\scriptscriptstyle C}}.
	\end{equation}
	The first and second terms of the rhs member respectively correspond to ${\cal L}^{\rm os}\rho_{\rm s}$ and ${\cal L}^{\rm oo}\rho_{\rm o}$ transitions. Thanks to identities (\ref{eq:L0}) and  (\ref{eq:L1}), the first one readily writes
	\beq
	\bra{{\rm o}^{\scriptscriptstyle C}}L^{{\scriptscriptstyle A}}(\y)\ket{s}\rho_{\rm s}(t)\bra{s}L^{{\scriptscriptstyle A \dagger}}(\y) \ket{{\rm o}^{\scriptscriptstyle C}} &=& \frac{1}{2N_c} \int_{v\boldsymbol{X},v'\boldsymbol{X'} } g_-\left(t-v,\boldsymbol{y-X}\right)g_-^*\left(t-v',\boldsymbol{y-X'}\right)\nn
	&\times&  U^\dagger_{\rm o} (v-t)P_{\boldsymbol{X}}U_{\text{s}}\left(v-t\right)\rho_{\rm s}(t) U^\dagger_{\rm s}\left(v'-t\right)P_{\boldsymbol{X}'}U_{\rm o}(v'-t)\nn
	&=& \frac{1}{2N_c} L_{[\rm os]} \rho_{\text{s}} L^\dagger_{[\rm os]}.
	\eeq
	To evaluate the second term, one has to deal --- according to Eq.~(\ref{eq:L2}) --- with three subterms involving the convolutions $g_-\otimes g_-$, $g_+\otimes g_+$ and $g_-\otimes g_+$, with the following colour factors:
	\begin{eqnarray}
		g_-\otimes g_- &:& \frac{1}{4}\sum_{{\scriptscriptstyle A},{\scriptscriptstyle D}} d^{\scriptscriptstyle DAC} 
		d^{\scriptscriptstyle DAC} = \frac{N_c^2-4}{4 N_c}\nn
		g_+\otimes g_+ &:& \frac{1}{4}\sum_{{\scriptscriptstyle A},{\scriptscriptstyle D}} f^{\scriptscriptstyle DAC} 
		f^{\scriptscriptstyle DAC} = \frac{N_c}{4}\nn
		g_+\otimes g_- &:& \frac{i}{4}\sum_{{\scriptscriptstyle A},{\scriptscriptstyle D}} f^{\scriptscriptstyle DAC} 
		d^{\scriptscriptstyle DAC} = 0,
	\end{eqnarray}
	where the identities (\ref{eq:colour-identities}) have been used. The full expression writes
	\begin{eqnarray}
		\sum_{\scriptscriptstyle D}\bra{{\rm o}^{\scriptscriptstyle C}}L^{{\scriptscriptstyle A}}(\y)\ket{{\rm o}^{\scriptscriptstyle D}}\rho_{\rm o}\bra{{\rm o}^{\scriptscriptstyle D}}L^{{\scriptscriptstyle A \dagger}}(\y) \ket{{\rm o}^{\scriptscriptstyle C}} &=& \frac{N_c}{4} \int_{v\boldsymbol{X},v'\boldsymbol{X'} } g_+\left(t-v,\boldsymbol{y-X}\right)g_+^*\left(t-v',\boldsymbol{y-X'}\right)\nn
		&\times&  U^\dagger_{\rm o} (v-t)P_{\boldsymbol{X}}U_{\rm o}\left(v-t\right)\rho_{\rm o}(t) U^\dagger_{\rm o}\left(v'-t\right)P_{\boldsymbol{X}'}U_{\rm o}(v'-t)+\nn
		&&
		\frac{N_c^2-4}{4N_c} \int_{v\boldsymbol{X},v'\boldsymbol{X'} } g_-\left(t-v,\boldsymbol{y-X}\right)g_-^*\left(t-v',\boldsymbol{y-X'}\right)\nn
		&\times&  U^\dagger_{\rm o} (v-t)P_{\boldsymbol{X}}U_{\rm o}\left(v-t\right)\rho_{\rm o}(t) U^\dagger_{\rm o}\left(v'-t\right)P_{\boldsymbol{X}'}U_{\rm o}(v'-t)\nn
		&=& \frac{N_c}{4} L^{(+)}_{[{\rm oo}]} \rho_{\rm o} L^{(+)\dagger}_{[{\rm oo}]} + \frac{N_c^2-4}{4 N_c} 
		L^{(-)}_{[{\rm oo}]} \rho_{\rm o} L^{(-)\dagger}_{[{\rm oo}]},
	\end{eqnarray}
	where the following new reduced operators have been defined:
	\beq
	L^{\scriptscriptstyle (\pm)}_{[\rm oo]}(\y)= \int_{v,\boldsymbol{X}} g_\pm\left(t-v,\boldsymbol{y-X}\right)U^\dagger_{\rm o}\left(v-t\right)P_{\boldsymbol{X}}U_{\rm o}(v-t).
	\eeq
	Note that the upper script $\pm$ is in correspondence with the subscript in $g_\pm$. The calculation of 
	$\bra{{\rm o}^{\scriptscriptstyle C}}
	\{L^{{\scriptscriptstyle A}\dagger}(\y) L^{\scriptscriptstyle A}(\y),\rho \}\ket{{\rm o}^{\scriptscriptstyle C}}$ 
	proceeds in a similar way. It involves both the $L^\dagger_{\rm [so]} L_{\rm [so]}$, the $L^{(+)\dagger}_{\rm [oo]} L^{(+)}_{\rm [oo]}$, and the $L^{(-)\dagger}_{\rm [oo]} L^{(-)}_{\rm [oo]}$ products.
	
	By combining the previous results, one can rewrite the Lindblad equation as a coupled system of two equations for $\rho_{\rm s}$ and $\rho_{\rm o}$. After some rearrangements of terms, we obtain
	\beq\label{eq:Lindbladsofinal}
	\frac{{\rm d}\rho_{\rm s}}{{\rm d}t}  &=& \cdots +C_F L_{[\rm so]}\, \rho_{\text{o}}\, L^\dagger_{[\rm so]} -\frac{C_F}{2}\left\{L^\dagger_{[\text{os}]} L_{[\text{os}]},\rho_{\rm s}\right\}\nn
	\frac{{\rm d}\rho_{\rm o}}{{\rm d}t} &=& \cdots + \frac{1}{2 N_c}  L_{[\rm os]} \,\rho_{\text{s}}\, L^\dagger_{[\rm os]}-\frac{1}{4N_c} \left\{ L^\dagger_{[\text{so}]} \,L_{[\text{so}]},\rho_{\rm o} \right\} +  \nn
	&& \frac{N_c^2-4}{4N_c}\left( L^{\scriptscriptstyle (-)}_{[\text{oo}]}\,\rho_{\rm o}\, L^{{\scriptscriptstyle (-)}\,\dagger}_{[\text{oo}]}-\frac{1}{2}\left\{L^{\scriptscriptstyle (-)\,\dagger}_{[\text{oo}]} \,L^{\scriptscriptstyle (-)}_{[\text{oo}]},\rho_{\rm o}\right\}\right)+\frac{N_c}{4}\left( L^{\scriptscriptstyle (+)}_{[\text{oo}]}\,\rho_{\rm o}\, L^{{\scriptscriptstyle (+)}\,\dagger}_{[\text{oo}]} -\frac{1}{2 }\left\{L^{\scriptscriptstyle(+)\dagger}_{[\text{oo}]} L^{\scriptscriptstyle(+)}_{[\text{oo}]},\rho_{\rm o}\right\}\right),\nn
	\eeq
	where ``$\cdots"$ represents the (temporarily omitted) contribution of the unitary terms. In addition, one easily shows that $\bra{o^{\scriptscriptstyle C}}\rho\ket{s}$ elements are not generated from the evolution using the relations $d^{\scriptscriptstyle ABC}\delta^{\scriptscriptstyle AB}=0$ and $f^{\scriptscriptstyle ABC}\delta^{\scriptscriptstyle AB}=0$, which would be unphysical. We note that, while the last line of the equation above for $\dot \rho_{\rm o}$ exhibits the expected Lindblad structure, this structure is not visible in the equation for $\dot \rho_{\rm s}$  and the first line of the equation for $\dot \rho_{\rm o}$. However this can be remedied by introducing  the following $2\times 2$ matrices
	\beq
	\rho=\left(\begin{array}{cc}
		\rho_{\rm s} & 0\\
		0 & \rho_{\rm o}
	\end{array}\right),\qquad \hat L_{[\text{os}]}= L_{[\text{os}]}\left(\begin{array}{cc}
		0 & 0\\
		1 & 0
	\end{array}\right),\qquad \hat L_{[\text{so}]}=L_{[\text{so}]}\left(\begin{array}{cc}
		0 & 1\\
		0 & 0
	\end{array}\right),
	\eeq 
	where we denote with a hat the operators acting in the two-dimensional space. It is then straightforward to verify that the first two lines of the equations above can be rewritten as 
	\begin{equation}
		\frac{{\rm d}\hat{\rho}}{{\rm d}t}= \left(\begin{array}{cc}
			C_F & 0\\
			0 & \frac{1}{2N_c}
		\end{array}\right) \left(  \hat L_{[\rm so]}\, \hat\rho\, \hat L^\dagger_{[\rm so]}-\frac{1}{2} \left\{ \hat L^\dagger_{[\text{so}]} \,\hat L_{[\text{so}]},\hat\rho \right\} + \hat L_{[\rm os]} \,\hat\rho\, \hat L^\dagger_{[\rm os]} -\frac{1}{2}\left\{\hat L^\dagger_{[\text{os}]} \hat L_{[\text{os}]},\hat\rho\right\}  \right),
	\end{equation}
	where the Lindblad structures are now evident. Matrices representations can also be introduced for the octet transitions in the last line of Eq.~(\ref{eq:Lindbladsofinal}). This is however trivial  since all the operators are then proportional to the projector on octets, namely \scalebox{0.6}{$  \left(\begin{array}{cc}
			0 & 0\\
			0 & 1
		\end{array}\right) $}.\\
	
	It remains to consider the unitary evolution and look at  the corrections $\Lambda$ to the heavy quark hamiltonian, $H\mapsto H_{\scriptscriptstyle Q}+\Lambda$, with $\Lambda$ given by Eq.~(\ref{eq:Lambda}). 
	The colour algebra is straightforward and uses some of the intermediate steps of the previous calculations.  In the matrix representation introduced above $\Lambda$ is diagonal and 
	\begin{equation}
		H_{\scriptscriptstyle Q}=\left(\begin{array}{cc}
			H_{\rm s} & 0\\
			0 & H_{\rm o}
		\end{array}\right),\qquad
		\Lambda=\left(\begin{array}{cc}
			\Lambda_{\rm s} & 0\\
			0 & \Lambda_{\rm o}
		\end{array}\right).
		\label{matrix-rho}
	\end{equation} For the singlet, we get
	\beq
	\Lambda_{\rm s}=\frac{C_F}{2\pi}P\! \int \frac{d\omega}{\omega}\int_\y \,L^\dagger_{[\text{os}]}(\omega,\y) L_{[\text{os}]}(\omega,\y), 
	\eeq
	and for the octet
	\beq
	\Lambda_{\rm o}=\frac{1}{2\pi}P\! \int \frac{d\omega}{\omega}\int_\y\left(\frac{N_c^2-4}{4 N_c}\ L^{\scriptscriptstyle (-)\,\dagger}_{[\text{oo}]} \,L^{\scriptscriptstyle (-)}_{[\text{oo}]}+\frac{N_c}{4} L^{\scriptscriptstyle(+)\dagger}_{[\text{oo}]} L^{\scriptscriptstyle(+)}_{[\text{oo}]} +\frac{1}{2N_c} L^{\dagger}_{[\text{so}]} L_{[\text{so}]} \right).
	\eeq
	
	In summary, the QCD Nathan-Rudner ULE can be written as follows :
	\beq \label{ULE-colour-projected}
	\frac{d\hat\rho\left(t\right)}{dt}&=&-i\left[H_{\scriptscriptstyle Q}+\Lambda, \hat\rho\left(t\right)\right]\nn 
	&+&\sum_{n}\gamma_n \int_{\boldsymbol{y}}\Bigl(\hat L_{n}\left(\boldsymbol{y}\right)\hat \rho\left(t\right)\hat L_{n}^{\dagger}\left(\boldsymbol{y}\right)
	-\frac{1}{2}\left\{ \hat L_{n}^{\dagger}\left(\boldsymbol{y}\right)\hat L_{n}\left(\boldsymbol{y}\right),\hat \rho\left(t\right)\right\} \Bigr), 
	\eeq 
	where the sum $\sum_{n}$ runs over the  set  $\left\{L_{[\rm so]},L_{[\rm os]},L^{(-)}_{[\rm oo]},L^{(+)}_{[\rm oo]}\right\}$ of Lindblad operators, and the matrices $\gamma_n$ take care of the  colour factors:
	\begin{equation}
		\gamma_{\rm{so}}=\gamma_{\rm{os}}=\left(\begin{array}{cc}
			C_F & 0\\
			0 & 1/2N_c
		\end{array}\right),\quad
		\gamma^{-}_{\rm{oo}}=\frac{N_c^2-4}{4N_c}\left(\begin{array}{cc}
			0 & 0\\
			0 & 1
		\end{array}\right), \quad\gamma^{+}_{\rm{oo}}=\frac{N_c}{4}\left(\begin{array}{cc}
			0 & 0\\
			0 & 1
		\end{array}\right).
	\end{equation}
	
	\subsection{Reduced Lindblad operators for the relative motion}
	
	One may want to focus on the study of the relative motion, and for that purpose define a reduced density matrix $\tilde \rho_{\rm s,o}$ obtained by tracing out the center of mass coordinates (or momentum), i.e. $\tilde \rho_{\rm s,o}=\int_\P \bra{\P} \rho_{\rm s,o} \ket{\P}$ with $\P$ the center of mass momentum. We briefly show here how to obtain the corresponding Lindblad operators acting on $\tilde \rho_{\rm s,o}$ (see \cite{blaizot2018approach} for more details). To avoid confusion we shall, in this section, denote by a tilde the operators which act on the relative coordinates.  
	
	We shall work out explicitly only the case of the operator ${\cal L}^{\rm ss}$ in Eq.(\ref{eq:Lindbladsofinal}), and look at one term of the anticommutator, namely 
	\beq\label{eq:PLP0}
	\int_\P \bra{\P}L^\dagger_{[\text{os}]} L_{[\text{os}]}\rho_{\rm s}\ket{\P}=\int_{\P,\P_1,\P_2} \bra{\P}L^\dagger_{[\text{os}]}\ket{\P_1}\bra{\P_1} L_{[\text{os}]}\ket{\P_2}\bra{\P_2}\rho_{\rm s}\ket{\P},
	\eeq 
	with $L_{[\text{os}]}$  given by Eq.~(\ref{eq:Los}), and we recall that an integration over $\y$ is involved in the product $L^\dagger_{[\text{os}]} L_{[\text{os}]}$. 
	We note that the evolution operators contained in $L_{[\text{os}]}$ is the product of two commuting operators  $U_{\rm s,o}(v)=U_{\rm cm}(v)\tilde{U}_{\rm s,o}(v)$, where $\bra{\P}U_{\rm cm}(v)\ket{\P}=\exp\left(-i v \frac{P^2}{4M}\right)$,   and $\tilde{U}_{\rm s,o}(v)$ acts on the relative coordinates.  
	We note  also that the coordinates $\boldsymbol{X}$ that are involved in the factor $g_-(t-v,\boldsymbol{y-X})$ are determined by the projector $P_{\boldsymbol{X}}$. It is then convenient to make these coordinates explicit, which can be achieved with a Fourier transform 
	\begin{equation}
		g(t-v,\boldsymbol{y}-\boldsymbol{x})=\int_{\q}\rme^{i\q\cdot (\y-\x)}g(t-v,\q)\label{eq:91},
	\end{equation}
	so that for instance
	\beq\label{eq:LosR}
	L_{[\rm os]}&=& \int_{v,\q,\x,\bar\x} g(t-v,\q)\bigl[\rme^{i\boldsymbol{q}\left(\boldsymbol{y}-\boldsymbol{x}\right)}-\rme^{i\boldsymbol{q}\left(\boldsymbol{y}-\bar{\boldsymbol{x}}\right)}\bigr] U^\dagger_{\rm o}(v-t)\ket{\x,\bar\x}\bra{\x,\bar\x}U_{\rm s}(v-t)\nn
	&=& -2i\int_{v,\q,\s,\R} g(t-v,\q)\,\rme^{i\q\cdot (\y- \R)}\,\sin\frac{\q\cdot\s}{2}\, U^\dagger_{\rm o}(v-t)\ket{\s,\R}\bra{\s,\R}U_{\rm s}(v-t),\nn
	\eeq
	where in the second line we have used relative and center of mass coordinates, $\s=\x-\bar\x$, and $\R=(\x+\bar\x)/2$. In order to calculate the matrix element $\bra{\P_1}L_{[\rm so]}\ket{\P_2}$
	we need to evaluate
	\beq
	\bra{\P_1}U^\dagger_{\rm cm}(v-t)\ket{\R}\bra{\R}U_{\rm cm}(v-t)\ket{\P_2}=\rme^{i(\P_2-\P_1)\cdot \R}\,\rme^{-i(v-t) \frac{P_2^2-P_1^2}{4M}}.
	\eeq
	The integration over $\R$ in Eq.~(\ref{eq:LosR}) then fixes  $\P_2=\P_1+\q$. By substituting this value of $\P_2$ in the time dependent phase factor we get 
	\beq
	\rme^{-i(v-t) \frac{P_2^2-P_1^2}{4M}}\mapsto\rme^{-i(v-t) \frac{(\P_1+\q)^2-P_1^2}{4M}}=\rme^{-i(v-t) \frac{\q^2 + 2\P_1\cdot\q}{4M}}
	\eeq 
	As argued in \cite{blaizot2018approach}, within the present approximation, this phase can be safely ignored. This is because, the $v$ integration is limited to $|v-t|\lesssim \tau_E\sim m_D^{-1}$ while $q\lesssim m_D$. It follows that $(v-t) \frac{q^2}{4M}\lesssim m_D/M\ll 1$. The same applies to the term involving $\P_1$ as long as $P_1\lesssim T$, which we assume here to be the case.
	We are then left with 
	\beq
	\bra{\P_1}L_{[\rm os]}\ket{\P_2}=-i\int_{v,\q} \delta (-\q+\P_2-\P_1) g(t-v,\q)\,\rme^{i\q\cdot \y}\,\tilde U^\dagger_{\rm o}(v-t){\cal S}_{\q\cdot\hat{\s}} \tilde U_{\rm s}(v-t)
	\label{eq:Los-decomp}
	\eeq
	where ${\cal S}_{\q\cdot\hat{\s}}\equiv 2 \sin \frac{\q\cdot \hat{\s}}{2}$, with $\hat{\s}$, the operator acting on relative coordinates. This matrix element can be written as
	\beq
	\bra{\P_1}L_{[\rm os]}\ket{\P_2}=\int_{\q} \delta (-\q+\P_2-\P_1) e^{i\q\cdot \y} \tilde L_{[\text{os}]}(\q) 
	\eeq
	where
	\beq\label{eq:Lsoq}
	\tilde L_{[\text{os}]}(\q)=-i\int_{v} g(t-v,\q)\, \tilde U^\dagger_{\rm o}(v-t) {\cal S}_{\q\cdot\hat{\s}} \tilde U_{\rm s}(v-t)
	\eeq
	acts on the relative coordinates. At this point we return to Eq.~(\ref{eq:PLP0}). We denote by $\q'$ the momentum conjugate to $\y$ in $L^\dagger_{[\text{os}]}$. The integration over $\y$ fixes $\q'=\q$, leading to
	\begin{eqnarray}
		\int_\P \bra{\P}L^\dagger_{[\text{os}]} L_{[\text{os}]}\rho_{\rm s}\ket{\P}&=&\int_{\q}
		\tilde L^\dagger_{[\text{os}]}(\q) 
		\tilde L_{[\text{os}]}(\q) 
		\int_{\P,\P_1,\P_2}
		\delta (-\q+\P-\P_1)
		\delta (-\q+\P_2-\P_1)\times
		\nn
		&&\hspace{4.5cm}\bra{\P_2}\rho_{\rm s}\ket{\P}
	\end{eqnarray}
	For any fixed $\q$, the product  of the delta functions $\delta (\q-\P+\P_1)\delta (-\q+\P_2-\P_1)$  yields $\P_1=\P-\q$, and $\P_2=\q+\P_1=\P$. Thus the matrix element (\ref{eq:PLP0}) can be written as follows 
	\beq
	\int_\q  \tilde L^\dagger_{[\text{os}]}(\q) 
	\tilde L_{[\text{os}]}(\q) \int_{\P}   \bra{\P}\rho_{\rm s}\ket{\P} =  \int_\q  \tilde L^\dagger_{[\text{os}]}(\q) 
	\tilde L_{[\text{os}]}(\q) \tilde{\rho}_{\rm s}
	\eeq 
	A similar strategy can be applied to all the Lindblad operators, paying attention to the fact that for the octet operators which involve $g_+$ in place of $g_-$, on should substitute $-i{\cal S}_{\q\cdot\hat{\s}}$ with ${\cal C}_{\q\cdot\hat{\s}}=2\cos \frac{\q\cdot\hat{\s}}{2}$.

	\section{Illustrations and contact with the quantum optical regime}
	\label{sub_sec_illustrations}
	
	While the solution of the QCD Nathan-Rudner equation established in the previous section is deferred to a future publication, an illustration of some of its key ingredients  are provided in this section. In particular we calculate  the dissociation rate of singlet bound states in terms of the Lindblad operators, as well as the spectral densities of singlet states, thereby making a brief contact with the phenomenology of quarkonia in a plasma.  We  also verify that the rates obtained with either the Redfield or the Nathan-Rudner equations nearly coincide, as they should, given the approximations involved in relating the two equations. We end this section with a brief remark on the quantum optical regime. 
	
	The singlet decay rates due to singlet-to-octet transitions are defined as ${\rm Tr}(\rho_{\rm s} \Gamma_{\rm s})$, with the operator\footnote{Since in this section we focus on the relative motion, we omit the tilde on corresponding operators to alleviate the notation.}
	\begin{equation}
		\Gamma_{\rm s} :=
		\frac{4 \pi \alpha_S C_F}{\hbar c} \int_{\q} {L}^\dagger_{[\rm{os}]}(\q)
		{L}_{[\rm{os}]}(\q),
	\end{equation}
	where ${L}_{[\rm{os}]}(\q)$ is the operator (\ref{eq:Lsoq}) and we have made explicit the dependence on the QCD coupling constant $\alpha_s$, while the factor $1/\hbar c$ ensures that the matrix elements of $\Gamma_{\rm s}$ has the dimension of an energy (GeV). The identification of $\Gamma_{\rm s}$ proceeds by noticing that the anticommutator term in  the first line of Lindblad Eq.~(\ref{eq:Lindbladsofinal}) can be used to define   an effective complex hamiltonian (see Eq.~(\ref{eq:Hseff}) below) of the generic form \cite{daley2014quantum}
	\beq
	\label{eq:Hseff}
	H_{\rm s}\mapsto H^{\rm eff}_{\rm s}: = H_{\rm s} -\frac{i}{2} \frac{4 \pi \alpha_S C_F}{\hbar c}\,L^\dagger_{[\text{os}]} L_{[\text{os}]}=H_{\rm s}-i\frac{\Gamma_{\rm s}}{2}.
	\eeq
	One can verify that the corresponding part of the  equation of motion for  the density matrix, $\dot \rho= -i \left(   H^{\rm eff}_{\rm s} \rho-\rho H^{\rm eff \, \dagger}_{\rm s} \right)$, reproduces the anticommutator in Eq.~(\ref{eq:Lindbladsofinal}).
	
	In order to calculate the matrix elements of $\Gamma_{\rm s}$ we introduce two basis in the $Q\bar Q$ Hilbert space, which correspond respectively to the basis which make $H_{\rm s}$ and $H_{\rm o}$ diagonal. We denote the former by $\{ \ket{n} \}$ and the latter by $\{ \ket{\k} \}$, and their corresponding eigenenergies by $E^{(\rm s)}_n$ and $E^{(\rm o)}_\k$. Note that the singlet channel may contain bound states, while all the  states of the octet channel are continuum states.  In order to calculate the matrix element $\langle m |{\Gamma}_{\rm s} |n\rangle$, we insert complete sets of states at appropriate places. We are then led to evaluate, for instance\footnote{In the defined basis, the elements of the other Lindblad operators write explicitly as 
		$\bra{n}L_{[\rm{so}]}\left(\q\right)\ket{\boldsymbol{k}}=-i g\left(E^{(\rm o)}_{\boldsymbol{k}}-E^{(\rm s)}_n,\boldsymbol{q}\right)
		\langle n| S_{\boldsymbol{q}\cdot\hat{\boldsymbol{s}}}| \boldsymbol{k}\rangle$,
		$\bra{\boldsymbol{k}}{L}_{[\rm{oo}]}^{(-)}\left(\q\right)\ket{\boldsymbol{k}^\prime}=-ig\left(E^{(\rm o)}_{\boldsymbol{k'}}-E^{(\rm o)}_{\boldsymbol{k}},\boldsymbol{q}\right)\langle \boldsymbol{k}| S_{\boldsymbol{q}\cdot\hat{\boldsymbol{s}}}|\boldsymbol{k}^\prime\rangle$, and
		$\bra{\boldsymbol{k}}L_{[\rm{oo}]}^{(+)}\left(\q\right)\ket{\boldsymbol{k}^\prime}=g\left(E^{(\rm o)}_{\boldsymbol{k'}}-E^{(\rm o)}_{\boldsymbol{k}},\boldsymbol{q}\right)\langle \boldsymbol{k}| C_{\boldsymbol{q}\cdot\hat{\boldsymbol{s}}}| \boldsymbol{k}^\prime\rangle$} 
	\beq\label{eq:Los33}
	\bra{\k} L_{[\text{os}]}(\q)\ket{n}&=&-i\int_{v} g(t-v,\q)\, \bra{\k} U^\dagger_{\rm o}(v-t) {\cal S}_{\q\cdot\hat{\s}} \tilde U_{\rm s}(v-t)\ket{n}\nn
	&=&-i\int_{v} g(t-v,\q)\, \rme^{i\left(E^{(\rm o)}_\k-E^{(\rm s)}_n\right)(v-t)} \bra{\k} S_{\q\cdot\hat{\s}} \ket{n}\nn
	&=& -i g(E^{(\rm s)}_n-E^{(\rm o)}_\k,\q)\bra{\k} S_{\q\cdot\hat{\s}} \ket{n}.
	\eeq
	
	Proceeding similarly for the matrix element $\bra{m} L^\dagger_{[\text{os}]}(\q)\ket{\k}$ one finally gets
	\begin{equation}
		\langle m |{\Gamma}_{\rm s} |n\rangle = \frac{4 \pi \alpha_S C_F}{\hbar c}
		\int_{\boldsymbol{k}}\int_{\boldsymbol{q}} {g}\left(E^{(\rm s)}_m-E^{(\rm o)}_{\boldsymbol{k}},\q\right)
		{g}\left(E^{(\rm s)}_n-E^{(\rm o)}_{\boldsymbol{k}},\q\right)
		\bra{m} S_{\q\cdot\hat{\s}}\ket{\k} \bra{\k} S_{\q\cdot\hat{\s}}\ket{n}.\label{eq:Gammamn2}
	\end{equation}
	For the Redfield equation derived   in \cite{blaizot2018approach} (see Eq. (2.38) therein), the corresponding   matrix element is 
	\begin{equation}
		\langle m |{\Gamma}^{\rm BE}_{\rm s} |n\rangle = \frac{4 \pi \alpha_S C_F}{\hbar c}
		\int_{\boldsymbol{k}}\int_{\boldsymbol{q}} \Delta^>\left(E^{(\rm s)}_n-E^{(\rm o)}_{\boldsymbol{k}},q\right)
		\bra{m} S_{\q\cdot\hat{\s}} \ket{\k}
		\bra{\k} S_{\q\cdot\hat{\s}} \ket{n}.
		\label{eq:decayrate-BE-Redfield}
	\end{equation}
	As an obvious consequence, the diagonal elements 
	$\bra{n}\Gamma_{\rm s}\ket{n}$ and $\bra{n}\Gamma_{\rm s}^{\rm BE}\ket{n}$ are identical. For the non-diagonal elements, $m\neq n$, the appearance of the two energy gaps in (\ref{eq:Gammamn2}) can be traced back to the two times $v$ and $v'$ stemming from the correlator decomposition into the product of jump correlators, while only a single energy gap is found in the Redfield case. The impact of this difference will be discussed below.
	
	In the numerical calculations presented in this section, the physical space for the relative distance $\boldsymbol{s}$ is restricted to a sphere of radius $R=10$ fm, imposing vanishing boundary conditions at the surface. We concentrate on the $c\bar{c}$ systems, with $M_c$ taken as 1.5 GeV. To evaluate the  singlet states, one uses the 3D Cornell-like real potential 
	\begin{equation}
		V_{\rm Cornell}(s)={\rm min}\left(c_1 -\frac{\alpha \hbar c}{s} 
		+\frac{\sigma}{\hbar c} s,V_{\rm max}\right),
		\label{eq:Cornell}
	\end{equation}
	with $c_1=-0.161\,{\rm GeV}$, $\alpha=0.513$ and $\sigma=(0.412\,{\rm GeV})^2$. This potential saturates at large values of $s$ to a constant value $ V_{\rm max}\approx 0.87$ GeV; it supports three bound states, whose masses are, respectively, 3.16 GeV ($J/\psi$), 3.56 GeV ($\chi_c$) and 3.72 GeV ($\psi'$).
	The potential in the octet channel is chosen to be constant, equal to $V_{\rm max}$ everywhere.  
	
	To obtain a model expression for the jump correlator ${g}\left(q_{0},\boldsymbol{q}\right)=\sqrt{\Delta^{>}\left(q_{0},\boldsymbol{q}\right)} $,  we use the Hard Thermal Loop (HTL) approximation  for $\Delta^{>}\left(q_{0},\boldsymbol{q}\right)$, which yields (see e.g. \cite{beraudo2008real})
	\begin{equation}
		{g}\left(q_{0},\boldsymbol{q}\right) =  \sqrt{(1+N\left(q_{0}\right))\,\sigma\left(q_{0},\boldsymbol{q}\right)}  \label{jump-spectral-link}
	\end{equation}
	where $\beta=\frac{1}{T}$, $N\left(q_{0}\right)=1/\left(e^{\beta q_{0}}-1\right)$, and $\sigma\left(q_{0},\boldsymbol{q}\right)$ is the associated spectral function.
	We recall that this quantity vanishes whenever $|q_0|>\|\q\|$.
	
	Because of the plasma assumed isotropy,  $g$ depends only on $\|\boldsymbol{k}\|$ and $\|\boldsymbol{q}\|$. It is then natural to use basis of states that carry good angular momentum. One obvious consequence of isotropy is that $\bra{m}\Gamma_{\rm s}\ket{n}$ has a  band-diagonal structure with blocks $\langle m{\rm S} |\Gamma_{\rm s} |n{\rm S}\rangle $, $\langle m{\rm P} |\Gamma_{\rm s} |n{\rm P}\rangle $\ldots for s,p,\ldots harmonics.  Each of the calculations in the present study implies a summation on spherical harmonics associated to the $|\boldsymbol{k}\rangle$ octet continuum states.  For compact singlet $|m\rangle$ and $|n\rangle$ states, the  
	$S_{\q\cdot\hat{\s}}$ operator induces transitions towards a limited range of these spherical harmonics, up to $l\sim \frac{\sqrt{\langle s^2\rangle} m_D}{\hbar c}$, where $\sqrt{\langle s^2\rangle}$ is the rms of the singlet states. For the continuum singlet states, the number of spherical harmonics to take into account in order to ensure convergence increases up to $\approx \frac{R m_D}{\hbar c} \approx 60$ for the highest temperatures considered in this study.
	
\begin{figure}[H]
		\centering
		\includegraphics[width=0.56\linewidth]{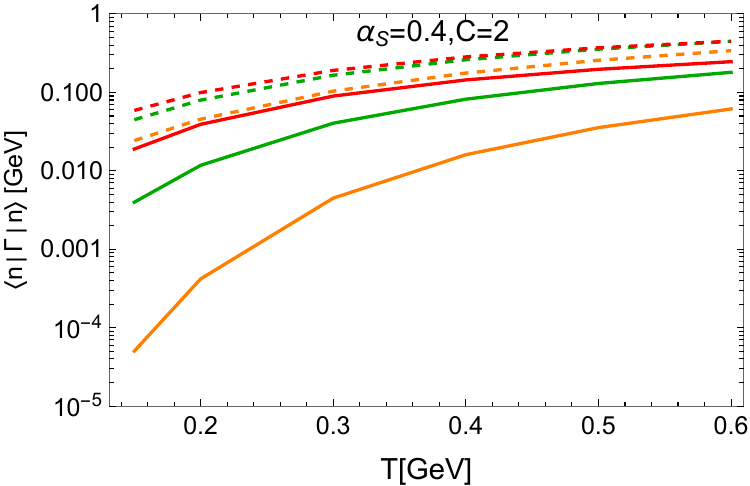}
		\includegraphics[width=0.41\linewidth]{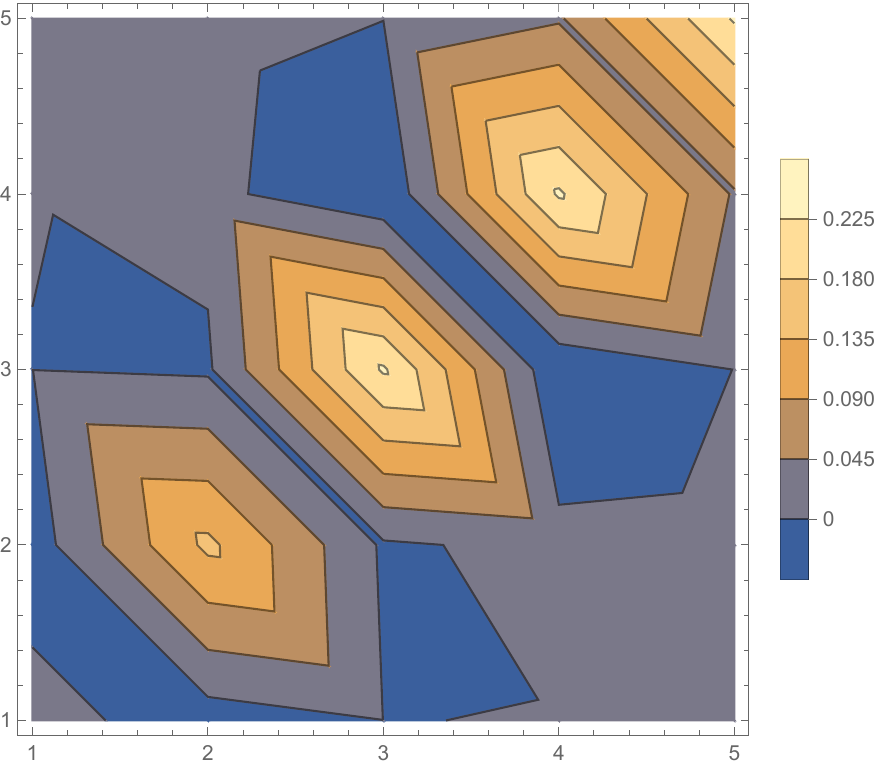}
		\caption{Left: Plain curves illustrate the diagonal elements 
			$\langle n|\Gamma_{\rm s}|n\rangle$ as a function of the QGP temperature $T$ for the bound $c\bar{c}$ vacuum-states: $J/\psi$ (orange), $\chi_c$ (green) and $\psi'$ (red); the correlator is evaluated with $m_D$ taken as $2T$ and $\alpha_s=0.4$. Dashed curves represent the same quantities neglecting the effect of energy gaps (imaginary potential approximation). Right: Contour plot of the elements $\bra{m}\Gamma_{\rm s}\ket{n}$  for $T=0.4$ GeV, and $n,m\le 5$.  Same choice for $\alpha_s$ and $m_D/T$ as in the left panel. }
		\label{fig:Gammamn3D}
	\end{figure}

	Figure \ref{fig:Gammamn3D} (plain lines) presents the decay rates of the 3 bound singlet-states that result from transitions to the octet representation. One observes a natural increase with $T$ as well as the expected hierarchy of excited states. In the same panel, the dashed curves illustrate the rates computed by setting the energy gaps $E_m^{(\rm s)}-E_\k^{(\rm o)}$ and $E_n^{(\rm s)}-E_\k^{(\rm o)}$ to 0 in the jump correlators, which would also result from the folding of the imaginary potential with the $\langle \boldsymbol{s}|n\rangle$ wave functions. The reduction of the rate due to the energy gaps is sizeable, especially for low $T$ and the deepest bound states, as expected. This effect was first discussed in \cite{Blaizot:2021xqa} and is common to both the QCD Nathan-Rudner and Redfield equations since, as already pointed out,  $\bra{n}\Gamma_{\rm s}\ket{n}=\bra{n}\Gamma_{\rm s}^{\rm BE}\ket{n}$.
	On the right panel of Fig.~\ref{fig:Gammamn3D}, the matrix elements 
	$\bra{m}\Gamma_{\rm s}\ket{n}$  are represented for $T=0.4$ GeV and different values of $n$ and $m$. One observes a dominance of the diagonal elements, implying that the non-unitary part of the evolution causes a limited mixing between the eigenstates of $H_{\scriptscriptstyle Q}$. This also indicates that the $H_{\scriptscriptstyle Q}$ eigenstates can  account for the high-$T$ QBM Regime (provided that sufficiently enough partial waves are taken into account). This is confirmed by the  observed saturation of the diagonal elements $\bra{n} \Gamma_{\rm s}\ket{n}$ with increasing $n$, which, as we have verified, can be matched --- for singlet diffusion states --- to the diffusion coefficient of two independent $c$ quarks. 
	
\begin{figure}[H]
		\centering
		\includegraphics[width=0.49\linewidth]{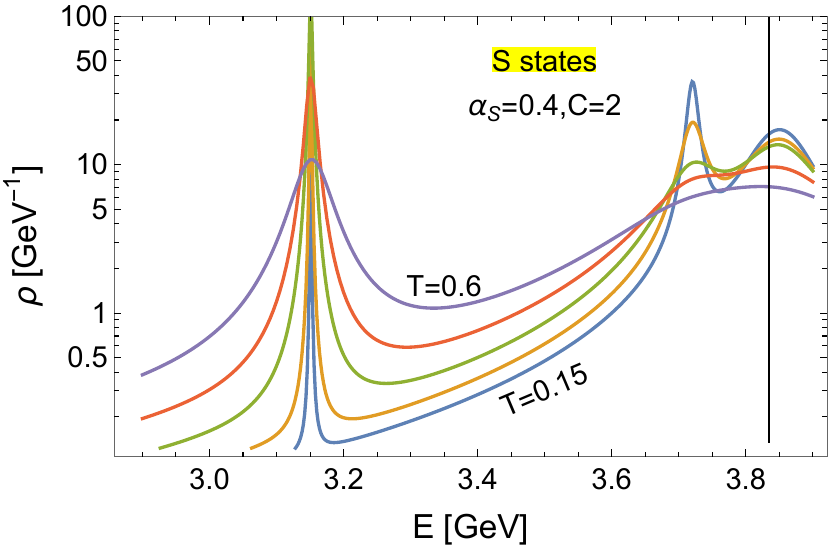}
		\includegraphics[width=0.49\linewidth]{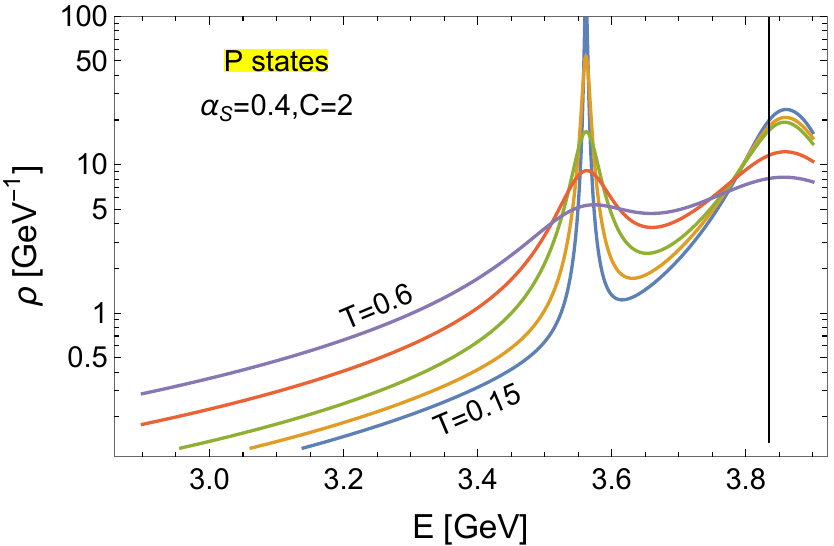}
		\caption{Left: Spectral density according to Eq.~(\ref{eq:spectral2}) for $T=0.15$, 0.2, 0.3, 0.4 and 0.6 GeV (S states). Right: same for P states.}
		\label{fig:spectraldensity3D}
	\end{figure}

	In order to gain further insight into the dissociation of in-QGP quarkonia bound states, we compute the normalised spectral  density defined as 
	\begin{equation}
		\rho_{\rm s}(E):=\frac{1}{\pi}\Im\left( {\rm tr}\frac{1}{{H}_{\rm s}^{\rm eff}-E {I}}\right).
		\label{eq:spectral2}
	\end{equation}
	Here, ${H}_{\rm s}^{\rm eff}$ is the effective non-unitary Hamiltonian introduced at the beginning of this section, 
	${H}_{\rm s}^{\rm eff}:={H}_{\rm s} - \frac{i}{2}{\Gamma}_{\rm s}$, where $H_{\rm s}$ is the sum of the kinetic hamiltonian and the real potential defined in  Eq.~(\ref{eq:Cornell}). In a first calculation, we neglect the effect of $\Lambda_{\rm s}$ in order to demonstrate the sole effect of the non-unitary part of the Lindblad equation. The corresponding spectral density  is plotted as a function of the energy $E$ in Fig.~\ref{fig:spectraldensity3D}. One observes a continuous transition from a structured  spectral density with multiple resonances to a structureless density when passing from low to high temperatures. Notice that the $\psi(2{\rm S})$ peak is still present for $T$ as high as 0.3 GeV while the $\psi(1{\rm S})$ peak survives for temperatures as high as 0.6 GeV. Such robustness partly stems from the fact that the modification of the potential at finite $T$ due to $\Lambda_{\rm s}$ is neglected in the calculation, which corresponds to evaluating the elements of ${\Gamma}_{\rm s}$ with vacuum-eigenstates at all temperatures.\footnote{It is worth to note that for the HTL description of the QGP, the off-diagonal elements of $\Gamma_{\rm s}$ in Eq. (\ref{eq:Gammamn2}) are found to play a very little role in the peak broadening and for the spectral function in general.} 
	
	\begin{figure}[H]
		\centering
		\includegraphics[width=0.49\linewidth]{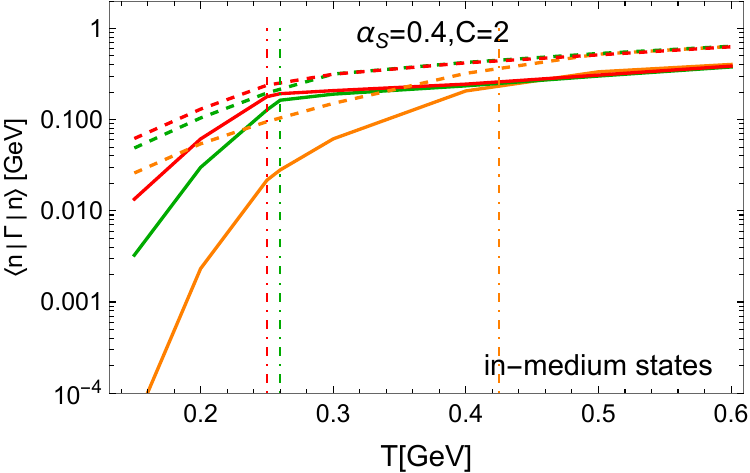}
		\includegraphics[width=0.49\linewidth]{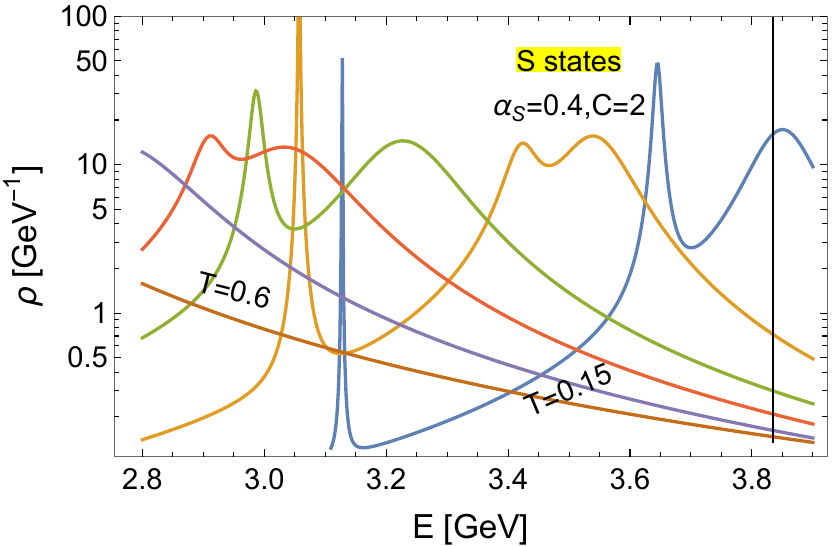}
		\caption{Left: same as Fig.~\ref{fig:Gammamn3Dinmed} (left) for in-medium eigenstates of the effective $T$-dependent real-potential defined in \cite{lafferty2020improved}; the dot-dashed vertical lines represent, for each state, the $T$ beyond which it ceases to be bound, namely $T_{\rm diss}(J/\psi)\approx 0.41\,{\rm GeV}$, $T_{\rm diss}(\chi_c)\approx 0.26\,{\rm GeV}$, and $T_{\rm diss}(\psi')\approx 0.25\,{\rm GeV}$. Right: same as Fig.~\ref{fig:spectraldensity3D} (left) for in-medium states, for $T=0.15$ (blue), 0.2 (orange), 0.25 (green), 0.3 (red), 0.4 (purple) and 0.6 (brown) GeV. 
		}
		\label{fig:Gammamn3Dinmed}
	\end{figure}
	In Fig.~\ref{fig:Gammamn3Dinmed}, we study the consequence of resorting to a $T$-dependent real potential in the calculation of the spectral density. As a proxy for $V_{\rm Cornell}+\Lambda_{\rm s}(T)$, we use the potential defined in \cite{lafferty2020improved}.
	As expected, the decay widths increase much faster than with the Cornell-like potential, until the various dissociation temperatures for which the states become unbound. Similarly, the corresponding spectral density loses its structure for lower temperature, as illustrated on the right panel of Fig.~\ref{fig:Gammamn3Dinmed}, which reproduces the qualitative features of spectral functions stemming for instance from lQCD \cite{Burnier:2015tda,Ding:2017std,Larsen:2026qvs}. 
	
	We now return to the issue of the different energy gaps that enter the jump correlators or the correlator $\Delta^>$ in the expressions of the singlet decay rates, respectively Eqs.~(\ref{eq:Gammamn2}) and (\ref{eq:decayrate-BE-Redfield}). As mentioned earlier, 
	while the Nathan-Rudner equation treats both associated energy gaps $E^{(\rm o)}_\k-E^{(\rm s)}_{n'}$ and $E^{(\rm o)}_{\k'}-E^{(\rm s)}_{n}$ symmetrically in the product of the jump correlators, the  Redfield equation only retains one of these energy gaps
	in $\Delta^>(E^{(\rm o)}_\k-E^{(\rm s)}_{n'})$. Hence, differences may appear when both transitions involve gaps much larger than $m_D$, which is the case when dealing with quantum coherences for which $n$ and $n'$ differ by large amounts. Note, however, that such contributions  lead to rapidly oscillatory terms that soon become subdominant compared to the common secular terms found for vanishing gaps (see e.g. appendix B of \cite{blaizot2018approach}). One may also anticipate effects when the energy gaps, without being very large,  are simply  of the order of $m_D$: This is the case in the optical regime which we briefly discuss now.  
	
\begin{figure}[H]
		\centering
		\includegraphics[width=0.49\linewidth]{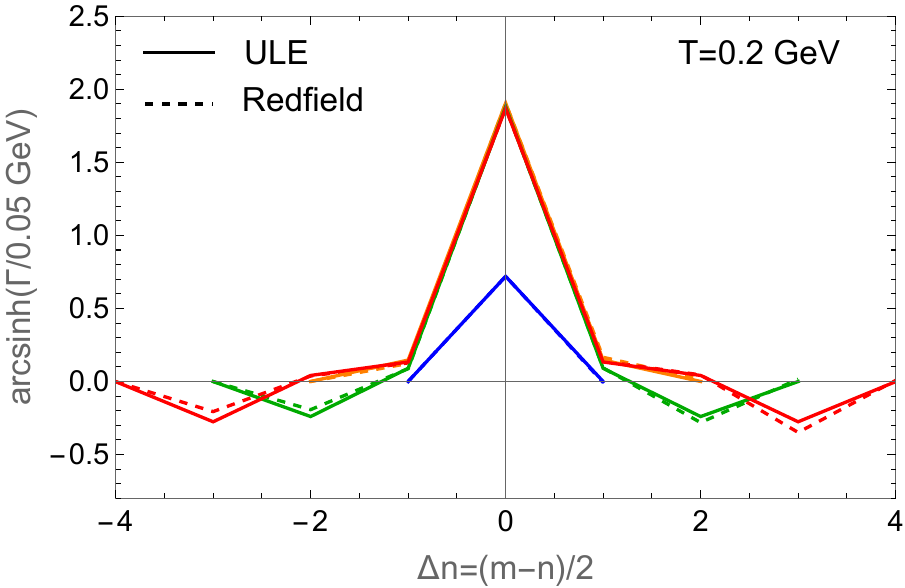}
		\includegraphics[width=0.49\linewidth]{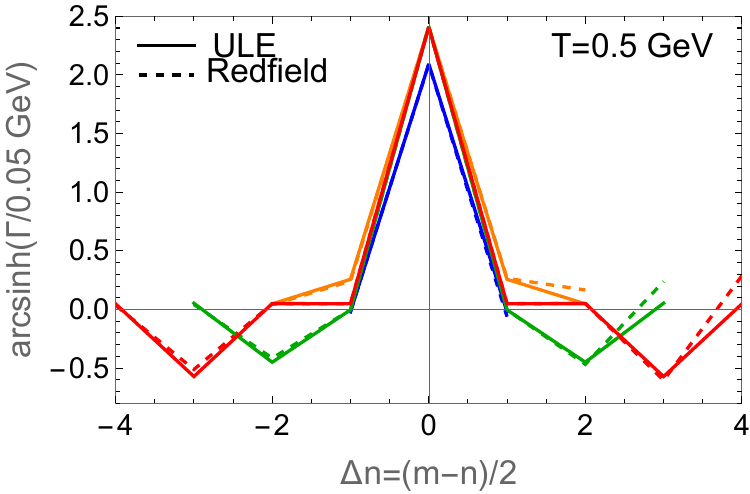}
		\caption{Left: Comparison between $\langle m |{\Gamma}_{\rm s}^{\rm BE} | n\rangle$ (dashed lines) and $\langle m |{\Gamma}_{\rm s}^{\rm ULE} | n\rangle$ (plain lines) in the s-channel for $m_D= 2 T$ and $T=0.2$ GeV. Each line correspond to given fixed sum $m+n$: 4 (blue), 6 (orange), 8 (green) and 10 (red); Using the arcsinh allows to compress the large values of $\langle m |{\Gamma}_{\rm s} | n\rangle$, making the small value more visible, while preserving the sign. Right: same for  $T=0.5$~GeV.}
		\label{fig:compGamma}
	\end{figure}

	The quantum optical regime is the regime where the Debye mass $m_D$ is of the order of the energy gap $\Delta E$, or equivalently the regime where $\tau_E\sim \tau_S$. To better understand the evolution of the system in this particular regime,  it is useful to return to the equation governing the singlet part of the density matrix (a similar argument applies to the octet  component).  By taking the matrix element between eigenstates of $H_{\rm s}$ of both sides of Eq.~(\ref{eq:singlet-equation-app}), and tracing out the center-of-mass, one gets
	\begin{equation}
		\frac{d\langle n|{\rho}_{\rm s}\left(t\right)|n^{\prime}\rangle }{dt}=-i\langle n|\left[{H}_{\rm s},{\rho_{\rm s}}\left(t\right)\right]|n^{\prime}\rangle +\langle n|{\mathcal{L}}_{\rm ss}\left(t\right){\rho_{\rm s}}\left(t\right)|n^{\prime}\rangle +\langle n|\mathcal{{L}}_{\rm so}\left(t\right){\rho}_{\rm o}\left(t\right)|n^{\prime}\rangle .
		\label{singlet-eq-projected-on-energy}
	\end{equation}
	The relevant matrix elements are given by 
	\begin{equation}
		\langle n|{\mathcal{L}}_{\rm ss}\left(t\right){\rho_{\rm s}}(t)|n^{\prime}\rangle  = -\frac{1}{2}\sum_m  \left( \bra{n}\Gamma_{\rm s}\ket{m} \bra{m} \rho_{\rm s}(t) \ket{n'} + \bra{n} \rho_{\rm s}(t) \ket{m}\bra{m} \Gamma_{\rm s}\ket{n} \right),
		\label{193}
	\end{equation}
	where $\bra{m}\Gamma_{\rm s}\ket{n}$ is given in Eq.~(\ref{eq:Gammamn2}), and similarly,
	\beq 
	\langle n|{\mathcal{L}}_{\rm so}\left(t\right){\rho_{\rm o}}\left(t\right)|n^{\prime}\rangle  & = & 4\pi \alpha_S C_{F}\int_{\q\k\k'} g\left(E^{(\rm o)}_\k-E^{(\rm s)}_{n'},\boldsymbol{q}\right)g\left(E^{(\rm o)}_{\k'}-E^{(\rm s)}_{n},\boldsymbol{q}\right)\nn
	&  & \times\langle n|S_{\boldsymbol{q}.\hat{\boldsymbol{s}}}|\boldsymbol{k}^{\prime}\rangle \langle \boldsymbol{k}^{\prime}|{\rho}_{\rm o}\left(t\right)|\boldsymbol{k}\rangle \langle \boldsymbol{k}|S_{\boldsymbol{q}.\hat{\boldsymbol{s}}}|n^{\prime}\rangle \label{194}.
	\eeq 
	For the Redfield equation of~\cite{blaizot2018approach} the matrix element of $\Gamma^{\rm BE}_{\rm s}$ has already been given in (\ref{eq:decayrate-BE-Redfield}). Note that all these matrix elements involve the non-diagonal matrix elements of the decay rate (such as for instance $\bra{m}\Gamma\ket{n}$). 
	Fig.~\ref{fig:compGamma} shows a  direct comparison of $\langle m |{\Gamma}_{\rm s} | n\rangle$ and $\langle m |{\Gamma}^{\rm BE}_{\rm s} | n\rangle$ as they appear in  Eq.~(\ref{193}). Differences are expected only for $m \neq n$. We can see that these differences are smaller towards  the region of small values of $m-n$ in Fig.~\ref{fig:compGamma}.  The differences remain overall quite small, even for relatively large values of $m-n$, showing that the Redfield and the QCD Nathan-Rudner  equations should generate solutions quite close in functional space. 
	
	Further insight is obtained in the QOR from the following heuristic argument. Let us look at the evolution for long times, $t_1-t_0\gg \tau_E,\tau_S$. To do so, it is useful to return to the interaction representation, whereby 
	$$\bra{n}\rho_{\rm s}(t) \ket{n'} = \bra{n}\bar{\rho}_{\rm s}(t) \ket{n'} e^{-i(E^{(\rm s)}_{n}-E^{(\rm s)}_{n'})t},$$
	with $\bar \rho $ is the density matrix in the interaction representation (see Sect.~\ref{sec:NathanRudner}). For simplicity we shall focus on the contribution of $\rho_{\rm o}$ alone (other contributions can be treated similarly). We get then
	\beq
	\frac{d\bra{n}\bar{\rho}_{\rm s}(t) \ket{n'} }{dt}&=&4\pi \alpha_S C_{F}\rme^{+i(E^{(\rm s)}_n-E^{(\rm s)}_{n'})t}\int_{\q\k\k'} g\left(E^{(\rm o)}_\k-E^{(\rm s)}_{n'},\boldsymbol{q}\right)g\left(E^{(\rm o)}_{\k'}-E^{(\rm s)}_{n},\boldsymbol{q}\right)\nn
	&\times &\int_{\q\k\k'} \bra{n} S_{\q\cdot\hat{\s}}\ket{\k'}
	\bra{\k}S_{\q\cdot\hat{\s}}\ket{n'} \rme^{-i(E^{(\rm o)}_{\k}-E^{(\rm o)}_{\k'})t}\bra{\k}\bar{\rho}_{\rm o}(t)\ket{\k}.
	\eeq
	The integration over time leads to the following integral 
	\beq
	\int_{t_0}^{t_1} \rmd t' \rme^{+i\bigl[ (E^{(\rm s)}_n-E^{(\rm o)}_{\k'})-(E^{(\rm s)}_{n'}-E^{(\rm o)}_{\k})\bigr]t'}\bra{\k}\bar{\rho}_{\rm o}(t')\ket{\k}.
	\eeq
	To the extent that we can ignore the slow time-dependence of $\bar{\rho}_{\rm o}(t')$ between $t_0$ and $t_1$, as commonly assumed in the QOR context~\cite{breuer2002theory}, this integral is proportional to $\frac{\sin \Delta E (t_1-t_0)}{\Delta E}$, where here $\Delta E\equiv (E^{(\rm s)}_n-E^{(\rm o)}_{\k'})-(E^{(\rm s)}_{n'}-E^{(\rm o)}_{\k})$. This  approaches $\delta(\Delta E)$ when $\Delta E (t_1-t_0)\sim (t_1-t_0)/\tau_S$ becomes large. In this limit the difference between the energy gaps are constrained to remain close to each other, and therefore also to the energy gap involved in the Redfield equation. 
	
	\section{Recovering the quantum Brownian motion regime}
	\label{sec_contact}
	
	In the present section, we explain how to recover from the QCD Nathan-Rudner equation (\ref{eq:ULEsch}), equations that are often used in the quantum Brownian motion regime.\footnote{See~\cite{blaizot2018quantum,akamatsu2022quarkonium,Delorme:2024rdo} for further details.} This regime occurs when the response of the plasma is fast compared to the characteristic time of the heavy quark motion, that is when 
	$\tau_E\ll \tau_S$. In this case we can approximate the time dependence of the heavy quark density as follows
	\beq 
	U^\dagger\left(v-t\right){n}^{\scriptscriptstyle A}(\x)U\left(v-t\right)\simeq {n}^{\scriptscriptstyle A}(\x) +(v-t)\dot{n}^{\scriptscriptstyle A}(\x), 
	\eeq
	with $\dot{n}^{\scriptscriptstyle A}(\boldsymbol{x})=i\left[H_{\scriptscriptstyle Q},n^{\scriptscriptstyle A}(\boldsymbol{x})\right]$. Once inserted in the expression of the Lindblad operator, one obtains
	\begin{equation}
		L^{\scriptscriptstyle A}\left(\boldsymbol{y}\right)\simeq L_{\rm \scriptscriptstyle QBM}^{\scriptscriptstyle A}\left(\boldsymbol{y}\right):=\int_{-\infty}^{+\infty}\text{d}v\int_{\boldsymbol{x}}g\left(t-v,\boldsymbol{y}-\boldsymbol{x}\right)\left[n^{\scriptscriptstyle A}(\boldsymbol{x})+\left(v-t\right)\dot{n}^{\scriptscriptstyle A}(\boldsymbol{x})\right],
		\label{eq:defLQBM}
	\end{equation}
	where the time difference $|v-t|$ is of the order of $\tau_E$. As a measure of the characteristic time $\tau_S$ is provided by the ratio $\dot n^{\scriptscriptstyle A}/n^{\scriptscriptstyle A}\sim 1/\tau_S$, the correction to the density $n^{\scriptscriptstyle A}(\boldsymbol{x})$ due to its time-dependence is indeed of order $\tau_E/\tau_S$ and small in the QBM regime.
	
	Since neither $n^{\scriptscriptstyle A}(\x)$ nor $\dot n^{\scriptscriptstyle A}(\x)$ depend on time variables, one can perform explicitly the integration over $v$ in the Lindblad operator. This is conveniently done using the Fourier transform (\ref{eq:10}) of the jump correlator.  This leads to an expression for the Lindblad operator that involves only the static ($q_0=0$) part of the jump correlator, viz.
	\begin{equation}
		L_{\rm \scriptscriptstyle QBM}^{\scriptscriptstyle A}\left(\boldsymbol{y}\right)=\int_{\boldsymbol{q}}\rme^{i\boldsymbol{q}\cdot\boldsymbol{y}}\left[
		g\left(q_0=0,\boldsymbol{q}\right) n^{\scriptscriptstyle A}(\boldsymbol{q})-
		i\left. \partial_{q_0} g\left(q_0=0,\boldsymbol{q}\right) \right|_{q_0=0}
		\dot{n}^{\scriptscriptstyle A}(\boldsymbol{q})\right].
	\end{equation}
	In order to simplify the writing we shall set $n^{\scriptscriptstyle A}(\q)\mapsto n^{\scriptscriptstyle A}_\bfq$, $g\left(q_0=0,\boldsymbol{q}\right)\mapsto g_\bfq$ and $\partial_{q_0} g\left(q_0=0,\boldsymbol{q}\right)\mapsto g'_\bfq$. Thus the hermitian comjugate of $L_{\rm \scriptscriptstyle QBM}^{\scriptscriptstyle A}\left(\boldsymbol{y}\right)$ reads
	\beq
	L_{\rm \scriptscriptstyle QBM}^{A \dagger}\left(\boldsymbol{y}\right)=\int_\bfq \rme^{-i\y\cdot\bfq}\left[ g_\bfq n^{\scriptscriptstyle A}_{-\bfq}-i g'_\bfq \dot n^{\scriptscriptstyle A}_{-\bfq}  \right],
	\eeq 
	where we have used the fact that $g(q_0,\bfq)$ is real and  that $n^{\scriptscriptstyle A}(\x)$ is hermitian, so that $n^{{\scriptscriptstyle A} \dagger}_\bfq=n^{\scriptscriptstyle A}_{-\bfq}$.

	By injecting these expressions of the Lindblad operators into Eq.~(\ref{eq:ULEsch}), one obtains a sum of terms quadratic in  $n^{\scriptscriptstyle A}$ and $\dot{n}^{\scriptscriptstyle A}$. Let us consider first the contribution of the term $L^{\scriptscriptstyle A} \rho L^{{\scriptscriptstyle A}\,\dagger}$, for which we get
	\beq
	\int_{\boldsymbol{y}}\,L^{{\scriptscriptstyle A}}\left(\boldsymbol{y}\right)\rho\left(t\right)L^{{\scriptscriptstyle A}\,\dagger}\left(\boldsymbol{y}\right)\longrightarrow \int_{\bfq}
	\Bigl[
	I_0\, n^{\scriptscriptstyle A}_\bfq \rho(t) n^{\scriptscriptstyle A}_{-\bfq} +  I_2\,\dot{n}^{\scriptscriptstyle A}_\bfq \rho(t) \dot{n}^{\scriptscriptstyle A}_{-\bfq} 
	\qquad\qquad\qquad \nn
	+  I_1 \, \bigl(
	n^{\scriptscriptstyle A}_{\bfq} \rho(t) 
	\dot{n}^{\scriptscriptstyle A}_{-\bfq} -
	\dot{n}^{\scriptscriptstyle A}_{\bfq} \rho(t) n^{\scriptscriptstyle A}_{-\bfq}\bigr) \Bigr].
	\label{eq:expansion1}
	\eeq
	Here,  $I_0,I_1,I_2$ are the following quantities
	\beq
	I_0=g_\bfq^2, \qquad I_1= i g'_\bfq \, g_\bfq ,\qquad I_2= (g'_\bfq )^2 .
	\label{jump-correlator-product-recoilless}
	\eeq 
	
	Clearly, the same quantities are involved in the term $\{ L^{{\scriptscriptstyle A}\,\dagger}L^{\scriptscriptstyle A},\rho\}$ of  Eq.~(\ref{eq:ULEsch})
	so that 
	\beq
	\int_{\boldsymbol{y}}\left\{ L^{{\scriptscriptstyle A}\,\dagger}\left(\boldsymbol{y}\right)L^{{\scriptscriptstyle A}}\left(\boldsymbol{y}\right),\rho(t)\right\}\longrightarrow \int_{\boldsymbol{x}\boldsymbol{x}^{\prime}}
	\Bigl[
	I_0\,  \{n^{\scriptscriptstyle A}_\bfq  n^{\scriptscriptstyle A}_{-\bfq},\rho(t) \}+I_2 \{\dot n^{\scriptscriptstyle A}_\bfq \dot n^{\scriptscriptstyle A}_{-\bfq},\rho(t) \}  \nn
	+  I_1\,  \Bigl(\{n^{\scriptscriptstyle A}_\bfq \dot n^{\scriptscriptstyle A}_{-\bfq}- \dot n^{\scriptscriptstyle A}\bfq  n^{\scriptscriptstyle A}_{-\bfq},\rho(t)\} \Bigr).
	\label{eq:expansion2}
	\eeq 
	
	It is worth emphasising that the  substitution of $L^{\scriptscriptstyle A}$ (and its complex conjugate) by $L^{\scriptscriptstyle A}_{\rm \scriptscriptstyle QBM}$ in Eq.~(\ref{eq:ULEsch})
	trivially preserves its Lindbladian structure, which would also clearly be the case if higher orders in the gradient expansion were to be considered. However, because the Lindblad operator $ L^{\scriptscriptstyle A}$ involve linear combinations of $ n_{\q}^{\scriptscriptstyle A}$  and  $\dot n_{\q}^{\scriptscriptstyle A}$ this structure is somewhat hidden in the  writings (\ref{eq:expansion1}, \ref{eq:expansion2}).\footnote{It becomes more obvious realising from the structure of $I_0$, $I_1$ and $I_2$ that $ n_{\q}^{\scriptscriptstyle A}$  and  $\dot n_{\q}^{\scriptscriptstyle A}$ only appear in these expressions through the linear combinations $g_{\q} n_{\q}^{\scriptscriptstyle A} + i g'_{\q} \dot n_{\q}^{\scriptscriptstyle A}$. } 
	
	Now, if the plasma is in thermal equilibrium, which we have been assuming all along, the quantities $I_0$, $I_1$ and $I_2$  obey simple relations that follow from the connection of the jump correlator with the static correlator $\Delta^>(q_0=0)$. By comparing $I_0$ with the Fourier transform of $\Delta^>$, Eq.~(\ref{eq:10}),  one readily identifies 
	\beq
	I_0=\Delta^>(q_0=0,\bfq).
	\eeq  
	It is interesting to note that even though the product of factors $g$ that appear in the products of Lindblad operators in the ULE are no longer in the form of the initial convolution (\ref{convolution-ULEa}) where they relate to the correlator $\Delta^>$, that relation is recovered in the static limit ($q_0=0$) implied by the QBM regime. The quantity $I_1$ can be evaluated similarly. A simple calculation, using $\Delta^>(q_0,\bfq)=[g(q_0,\bfq)]^2$, yields 
	\beq
	I_1=\left.\frac{i}{2} \frac{\partial}{\partial q_0} \Delta^>(q_0,\bfq)\right|_{q_0=0}.
	\label{eq:I1fromcorrel}
	\eeq
	Now, under general conditions, (KMS boundary conditions on the correlator $\Delta$ and the assumption that the  spectral function is an odd function of the energy variable (see e.g. \cite{Blaizot:2015hya}), we have (with $\beta=1/T$)
	\beq
	\left.\frac{\partial}{\partial q_0} \Delta^>(q_0,\bfq)\right|_{q_0=0}=\frac{\beta}{2} \Delta^>(q_0=0,\bfq).
	\label{eq:gprimeg}
	\eeq
	From this relation, and using again $\Delta^>(q_0,\q)=g^2(q_0,\q)$, one deduces that 
	\beq
	\left.\frac{\partial}{\partial q_0} g(q_0,\q)\right|_{q_0=0}=\frac{\beta}{4} g(q_0=0,\q).
	\eeq 
	These relations imply that $I_1$ and $I_2$ are simple multiples of $I_0$. Further progress is achieved by relating the static correlator to the so-called imaginary potential, $W(\bfq)=-\Delta^>(q_0=0,\bfq)$ (see e.g. \cite{Blaizot:2015hya}).  
	We have then
	\begin{equation}\label{eq:I_n-2}
		I_0=-W(\bfq),\qquad
		I_1=-\frac{i}{4T}W(\bfq),\qquad
		I_2=-\frac{1}{16T^2}W(\bfq).
	\end{equation}
	
	Consider now the correction to $H_{\scriptscriptstyle Q}$. This is given by Eq.~(\ref{eq:Lambda}), which in the present context reduces to 
	\beq
	\Lambda&=&\frac{1}{2\pi}P\int \frac{\rmd \omega}{\omega}\int_{\x \x' \y} g(-\omega, \x'-\y) g(-\omega,\y-\x) n^{\scriptscriptstyle A}(\x') n^{\scriptscriptstyle A}(\x)\nn
	&=& \frac{1}{2\pi}P\int \frac{\rmd \omega}{\omega}\int_{\q} \Delta^>(-\omega, \q)  n^{\scriptscriptstyle A}_{\q} n^{\scriptscriptstyle A}_{-\q},
	\eeq
	which acts as a modification of the real potential in $H_{\scriptscriptstyle Q}$. 
	
	By collecting all the previous results, one can then write the Lindblad equation, in the QBM regime, entirely in terms of  the imaginary potential $W(\bfq)$ and the correction $\Lambda$ to the real potential. It reads
	\begin{equation}
		\frac{d\rho\left(t\right)}{dt} =  -i\left[H_{\scriptscriptstyle Q}+\Lambda,\rho\left(t\right)\right]-\int_{\boldsymbol{x}\boldsymbol{x}^{\prime}}W(\bfq)\left(\bar{\bar n}^{\scriptscriptstyle A}_\bfq\, \rho\, \left(t\right)\bar{\bar n}^{{\scriptscriptstyle A}\dagger}_\bfq-\frac{1}{2}\left\{\bar{\bar n}^{{\scriptscriptstyle A}\dagger}_\bfq \bar{\bar n}^{\scriptscriptstyle A}_\bfq,\,\rho\left(t\right)\right\}\right),
		\label{8.46-2}
	\end{equation}
	with 
	\begin{equation}
		\bar{\bar{n}}^{{\scriptscriptstyle A}}_\bfq :=n^{{\scriptscriptstyle A}}_\bfq+ \frac{i}{4T}\dot{n}^{{\scriptscriptstyle A}}_\bfq.
		\label{def:nbar_QBM}
	\end{equation}
	Interestingly, a new Lindblad structure has emerged, with new Lindbald operators defined as  $\sqrt{\Delta(q_0=0,\bfq)}\, \bar{\bar n}^{\scriptscriptstyle A}_\bfq$. 
	The equation (\ref{8.46-2}) is in agreement with the QME that had been obtained e.g. in \cite{blaizot2018quantum}, by exploiting a freedom in the choice of time discretisation (see Appendix B of \cite{blaizot2018quantum}).
	
	\section{Conclusions and perspectives}
	\label{section:Conclusions}
	
	In this work, we have applied the universal Lindblad equation (ULE) introduced by Nathan and Rudner~\cite{nathan2020universal} to the evolution of  quarkonia in a quark-gluon plasma. The main steps of the derivation are presented in Sec.~\ref{NRQCD-ULEb}. Projection onto the colour representations yields a coupled set of singlet-octet equations written in explicit Lindblad form. Unlike previous formulations, these equations do not rely on any specific hierarchy between the plasma temperature and the characteristic quarkonium energy gaps, while their Lindbladian structure  guarantees the positivity of the evolution of the reduced density matrix. Furthermore, their structure enables a stochastic unravelling, for instance through the method of quantum trajectories~\cite{omar2022qtraj}, thereby potentially reducing the computational cost of future numerical simulations.
	
	To illustrate the physical content of these new equations, we have presented in Sec.~\ref{sub_sec_illustrations} some results obtained within this framework. In particular, we investigate the temperature dependence of the singlet decay rate induced by singlet-to-octet transitions for several quarkonium states, recovering the expected qualitative behaviours. We also analyse the spectral density associated with an effective non-hermitian Hamiltonian that includes the singlet decay operator $\Gamma_{\rm s}$, over a temperature-range spanning both the quantum Brownian and quantum optical regimes. The resulting spectral density evolves smoothly across both regimes, illustrating that the QCD Nathan Rudner equation provides a unified and physically consistent description of quarkonium spectral properties. Finally, we compare off-diagonal elements of the decay operator $\Gamma_{\rm s}$ stemming from the ULE with those obtained from the Redfield equation and find only negligible differences. A simple heuristic argument is presented to explain why this should be so in the quantum optical regime. 
	
	In Sec.~\ref{sec_contact}, we have explored the connection, in the quantum Brownian regime, between the QCD Nathan Rudner equation and master equations previously derived in the literature. We show that these equations, usually derived from the Redfield equation, naturally emerge as a limiting case of the universal Lindblad equation, thereby confirming the general conclusion that both the Redfield equation and the ULE generate solutions which are close in functional space, the later being however completely positive.
	
	The present overall analysis provides a unified perspective on existing formulations of open quantum systems for in-medium quarkonium dynamics, which were originally developed under different physical assumptions for low and high QGP-temperatures
	
	A natural continuation of the present work will be to solve the coupled singlet-octet universal Lindblad equations. This will enable a complete numerical study of in-medium quarkonium dynamics throughout the evolution of the QGP, including the continuous evolution from the quantum Brownian to the quantum optical regime as the medium cools down.\footnote{While performing this work, we have been made aware of a possible alternate strategy developed by the TUM group to deal with this situation sticking to non-Lindblad Redfield-like master equations (see https://indico.mitp.uni-mainz.de/event/436/timetable/\#20260421).} In this respect, it is interesting to remark that the four Lindblad operators $L_{\rm [so]}$, $L_{\rm [os]}$ and $L^{(\pm)}_{\rm [oo]}$ that we have identified are time-independent, which will facilitate the numerical implementation. 
	
	An equally important direction concerns the approach to thermal equilibrium and the equilibrium state predicted by the universal Lindblad framework, which remain largely unexplored for the quarkonium-QGP system. Generalisation of the universal Lindblad equation to other systems such as $QQQ$ or $2(\bar{Q}Q)$ could be interesting to address the production of other heavy hadrons in URHIC.    
	
	Finally, the contact made between the ULE and the QBM regime at Sec.~4 offers interesting perspectives. Indeed, it has been argued recently \cite{Scheihing-Hitschfeld:2023tuz} that the spectral density could develop at the NLO a contribution even in the energy variable. In such a  situation, Eq.~(\ref{eq:gprimeg}) is not valid anymore. This is may also be the case when the plasma is not in equilibrium so that the KMS relation does not hold, which occurs for instance in the early stage of a nucleus-nucleus collision (see e.g.~\cite{Avramescu:2023qvv,Pooja:2024rnn}). It would be interesting to see how the techniques developed in this paper could be generalised to such cases. 
	
	\acknowledgments
	The authors acknowledge fruitful discussions with Nora Brambilla, Tom Magorsch, Antonio Vairo, Miguel Escobedo, and Gabriela Barenboim. We are grateful to the Institute for Nuclear Theory of the University of Washington and the Mainz Institute for Theoretical Physics for their generous support in the organisation of recent events (https://www.int.was\ hington.edu/programs-and-workshops/25-3b, https://indico.mitp.uni-mainz.de/event/434, https://indico.mitp.uni-mainz.de/event/436) where this work has been presented and consolidated thanks to the interactions with the participants.
	
	\bibliographystyle{unsrt}
	\bibliography{biblio}
\end{document}